\documentclass[10pt,preprint]{aastex}

\usepackage{epsfig}
\usepackage{graphicx,natbib}
\usepackage{natbib}
\usepackage{amsmath,amssymb}

\def\micron{\mu m}

\def\mjup{ M_{\rm J}}

\def\beq{\begin{equation}}
\def\eeq{\end{equation}}

\shorttitle {Lyot project survey statistical analysis}
\shortauthors {Leconte and Soummer}

\begin{document}

\title {The Lyot Project Direct Imaging Survey of Substellar Companions: \\
Statistical Analysis and Information from Nondetections.}
\slugcomment{Draft version \today}


 \author{J\'er\'emy Leconte}
\affil{\'Ecole Normale Supérieure de Lyon, Centre de Recherche astrophysique de Lyon, Universit\'e de Lyon, UMR 5574 CNRS, 46 all\'ee d'Italie, 69364 Lyon Cedex 07, France}
\email{jeremy.leconte@ens-lyon.fr}
\author{R\'emi Soummer}
\affil{Space Telescope Science Institute, 3700 San Martin Drive, Baltimore MD 21218, USA}
\email{soummer@stsci.edu}
\author{ Sasha Hinkley}
\affil{Department of Astronomy
California Institute of Technology
1200 East California Boulevard.
Mail Code 249-17
Pasadena, CA 91125, USA.}
\email{shinkley@astro.caltech.edu}
\author{ Ben R. Oppenheimer, Anand Sivaramakrishnan, Douglas Brenner}
\affil{Astrophysics Department, American Museum of Natural History, Central Park West at 79th Street, New York, NY 10024, USA}
\email{bro@amnh.org,anand@amnh.org,dbrenner@amnh.org}
\author{Jeffrey Kuhn}
\affil{Institute for Astronomy, University of Hawaii, 2680 Woodlawn Drive, Honolulu, HI 96822, USA}
\email{kuhn@ifa.hawaii.edu}
\author{James P. Lloyd}
\affil{Department of Astronomy, Cornell University, Ithaca, NY 14853, USA}
\email{jpl@astro.cornell.edu}
\author{Marshall D. Perrin}
\affil{UCLA Department of Astronomy, Los Angeles, CA 90095, USA }
\email{mperrin@ucla.edu}
\author{Russell Makidon$\dagger$\footnote{Russell Makidon died on 2009 June 22. The other authors would like to dedicate this paper to his memory.}}
\affil{Space Telescope Science Institute, 3700 San Martin Drive, Baltimore MD 21218, USA}
\author{Lewis C. Roberts, Jr}
\affil{Jet Propulsion Laboratory, California Institute of Technology, 4800 Oak Grove Drive, Pasadena CA 91109, USA}
\email{lewis.c.roberts@jpl.nasa.gov}
\author{James R. Graham}
\affil{(Department of Astronomy, 601 Campbell Hall, University of California Berkeley, CA 94720, USA}
\email{jrg@berkeley.edu}
\author{Michal Simon}
\affil{Deptartment of Physics and Astronomy Stony Brook University, Stony Brook, NY 11794-3800, USA}
\email{michal.simon@sunysb.edu}
\author{Robert A. Brown}
\affil{Space Telescope Science Institute, 3700 San Martin Drive, Baltimore MD 21218, USA}
\email{rbrown@stsci.edu}
\author{ Neil Zimmerman}
\affil{Astrophysics Department, American Museum of Natural History, Central Park West at 79th Street, New York, NY 10024, USA}
\email{nzimmerman@amnh.org}
\author{Gilles Chabrier, Isabelle Baraffe}
\affil{ \'Ecole Normale Supérieure de Lyon, Centre de Recherche astrophysique de Lyon, Universit\'e de Lyon, UMR 5574 CNRS, 46 all\'ee d'Italie, 69364 Lyon Cedex 07, France}
\email{gilles.chabrier@ens-lyon.fr,ibaraffe@ens-lyon.fr}

\begin{abstract}
The Lyot project used an optimized Lyot coronagraph with Extreme Adaptive Optics at the 3.63m Advanced Electro-Optical System telescope (AEOS) to observe 86 stars from 2004 to 2007. 

In this paper we give an overview of the survey results and a statistical analysis of the observed nondetections around 58 of our targets to place constraints on the population of substellar companions to nearby stars.

The observations did not detect any companion in the substellar regime. Since null results can be as important as detections, we analyzed each observation to determine the characteristics of the companions that can be ruled out. For this purpose we use a Monte Carlo approach to produce artificial companions, and determine their detectability by comparison with the sensitivity curve for each star. All the non-detection results are combined using a Bayesian approach and we provide upper limits on the population of giant exoplanets and brown dwarfs for this sample of stars. Our nondetections confirm the rarity of brown dwarfs around solar-like stars and we constrain the frequency of massive substellar companions ($M>40\mjup$) at orbital separation between and 10 and 50 AU to be $\lesssim 20\%$.

\end{abstract}


\section{Introduction}\label{sect:intro}

The discovery over the past 15 years of hundreds of exoplanets orbiting nearby stars has launched the new and thriving field of exoplanetary science \citep{MBF05,US07}.  The vast majority of these detections have been achieved indirectly through the radial velocity and transit methods, revealing a population of planets very close-in to their host star.  These efforts, although very sensitive and highly prolific, are intrinsically biased toward sensing planets at small angular separations \citep{CBM08}.  Direct imaging surveys (e.g. \citealt{OGK01,SOH07,LDM07,NCB07}), on the other hand, will be sensitive to a significantly different class of planet and allow astronomers to extend our knowledge to planets out of reach to the radial velocity and transit techniques.  For example, correlations between the frequency of exoplanets with the metallicity of the host star \citep{FV05}, as well as correlations with the stellar mass \citep{JBM07} can be explored by targeting a broad range of stellar types with direct imaging.  Moreover, this technique will even allow observers to address the dependence of planetary frequency with the age, luminosity, and location (i.e.  cluster, field, halo population etc.) of the stellar system.
When direct imaging surveys are fully mature, astronomers will be able to address questions related to other characteristics of the planetary system. The architecture of planetary systems, i.e. the distribution of planetary-mass companions as a function of the semi-major axis, SMA $\gtrsim 5$ AU can readily be addressed through direct imaging of nearby stars.  This question has been explored extensively through numerical work (e.g. \citealt{RF96,JT08,SM09}), but needs to be further constrained by high-contrast imaging survey results.  Similarly, the mass distribution of known exoplanets shows a power-law type behavior suggesting a large number of low-mass ($< 5 \mjup$) planets are yet to be discovered \citep{MBF05}.  Further, the shape of the companion mass function, and the issue of its universality, have already started to be explored through high-contrast imaging surveys \citep{MH09}.

In addition to easing some of the biases of the radial velocity and transit surveys, direct imaging will facilitate the detailed study of individual exoplanets and brown dwarfs.  The physics of these cool objects is highly complex and their atmospheric chemistry dominates their luminosity and evolution \citep{BCB03}.  Direct imaging opens the door to the direct study of these objects either through spectral analysis, or through broadband photometry combined with detailed evolutionary models.  Recent high contrast imaging detections of planetary mass companions \citep{MMB08,KGC08} have demonstrated the power of this technique and its potential for characterization of those objects out of reach to other methods.

The major obstacle to direct imaging of exoplanets and brown dwarfs, of course, is the overwhelming brightness of the host star.  A promising method for direct imaging of faint companions involves two techniques working in conjunction.  First, high-order adaptive optics (AO); provides control and manipulation of the image by correcting the aberrations in the starlight's wave front caused by Earth's atmosphere.  Second, a Lyot coronagraph \citep{Lyo39,SKM01} suppresses this corrected light.  In this work, we present an analysis of a direct imaging survey at the AEOS telescope on Haleakala, Hawaii \citep{ODN04,SOH07}.  After giving an overview of the instrument and Project (Section 2.1 and 2.2), we discuss challenges specific to high-contrast imaging projects (Section 2.3), namely the highly persistent speckle noise \citep{SFA07,HOS07} that limits the high-contrast sensitivity.  Next, we describe our technique to determine the survey sensitivity, based on a signal-to-noise (S/N) calculation (Section 3). We conclude with a detailed statistical analysis including 58 stars which is used to evaluate the completeness of our study and the significance of our lack of detections.

\section{Observations and Data processing}\label{dataprocess}

\subsection{Instrument}

The Lyot project \citep{ODN04,SOH07} employed an optimized, diffraction-limited Lyot coronagraph \citep{Lyo39,Mal96,SKM01} coupled to an infrared camera. It was deployed at the 
$D=3.63$ m Advanced Electro-Optical System (AEOS) telescope in Maui, 
with an AO system equipped with a 941 actuator deformable mirror (DM). 
The telescope is an altitude-azimuth design, with a beam traveling to a Coud\'e room containing the AO system \citep{RN02}.
The AO system's 941 actuator DM is complemented by a tip/tilt loop 
capable of running up to $\sim 4$ kHz, a Shack-Hartmann wave front sensor with 
a 2.5 kHz frame rate, and a real-time wave front reconstructor using least-square calculations performed on dedicated hardware \citep{RN02}. This combination of features can deliver some of the highest-order correction in modern AO.

The Lyot project coronagraph, housed in one of the Coud\'e rooms, received a 10cm collimated beam from the AO system.  Within the coronagraph a second-stage fast steering mirror (FSM) provided precision alignment with 3 mas resolution and no significant jitter of the star as demonstrated by \citet{DHO06}. The FSM was mounted on a piezo-electric  actuated tip/tilt stage, with the goal of  maintaining the PSF core on the center of the focal plane mask (FPM).  Rather than using an opaque mask as an occulter, the coronagraphic mask was comprised of a 455 $\mu$m hole drilled into a gold-coated reflective surface with $\lambda/20$ rms intrinsic surface irregularity. The coronagraphic suppression was realized by light falling through the FPM hole, and the outer portions of the image being reflected onward to the rest of the optical path. The instrument as a 2'' outer working angle (OWA). In addition, the light falling through the FPM fed an array of lenslets in a ``quad-cell'' configuration.  This quad-cell array, in turn, fed a set of four avalanche photodiodes, which were read at 2 kHz, and communicated tip tilt information back to the FSM via a centroiding algorithm. In addition, the coronagraph had a pupil-viewing CCD camera collecting data at a video rate.  This camera was purely for purposes of pupil alignment since the efficacy of the coronagraphic suppression depends heavily on the pupil alignment at the Lyot stop.  
The Lyot coronagraph had the capacity for simultaneous dual-beam polarimetry, achieved through the use of a Wollaston prism and liquid-lrystal variable retarders with the goal of obtaining Stokes $I$, $Q$, $U$ and $V$ images. There has been successive implementation of the polarimetric mode in the Lyot project, and details on the polarimetric modes are described elsewhere \citep{OBH08,HOS09}.  The present study considers the $I$ image only.
 
The primary science camera used in the Lyot project survey was the ``Kermit'' infrared camera \citep{PGT03} which is built around a Rockwell Hawaii-2 2048$\times$2048 HgCdTe detector. The camera optics utilize a unit-magnification, telecentric Offner relay, as well as a cold pupil stop within the camera dewar to reduce thermal background emission. The camera achieves a pixel scale of 13.51 mas/pixel. This pixel scale has been calculated by measuring the pixel separations of several binary stars with well-calibrated orbits. The Kermit readout electronics are based on hardware from Astronomical Research Cameras, Inc. and software adapted from that used by the Lick Observatory infrared camera, IRCAL \citep{LLM00}. 

\subsection{Survey description and observing strategy}

In 2004 March, the project began a survey of 86 nearby stars in coronagraphic mode in the $H$ band with the goal of detecting faint companions and disks orbiting the stars. Some targets were re-observed in $J$ and $K_s$ bands but these images are not included in this analysis.
The initial target list included the comprehensive list of all stars within 25pc - within the observable declination range at AEOS - that could be used as AEOS AO guide stars ($V\lesssim7$). This list was then prioritized into smaller categories: those with known disks, those with known exoplanets, those younger than 1 Gyr, and the remaining stars within 25pc that met our criteria. 

\begin{figure}[htbp]
\center
\resizebox{.4\hsize}{!}{\includegraphics {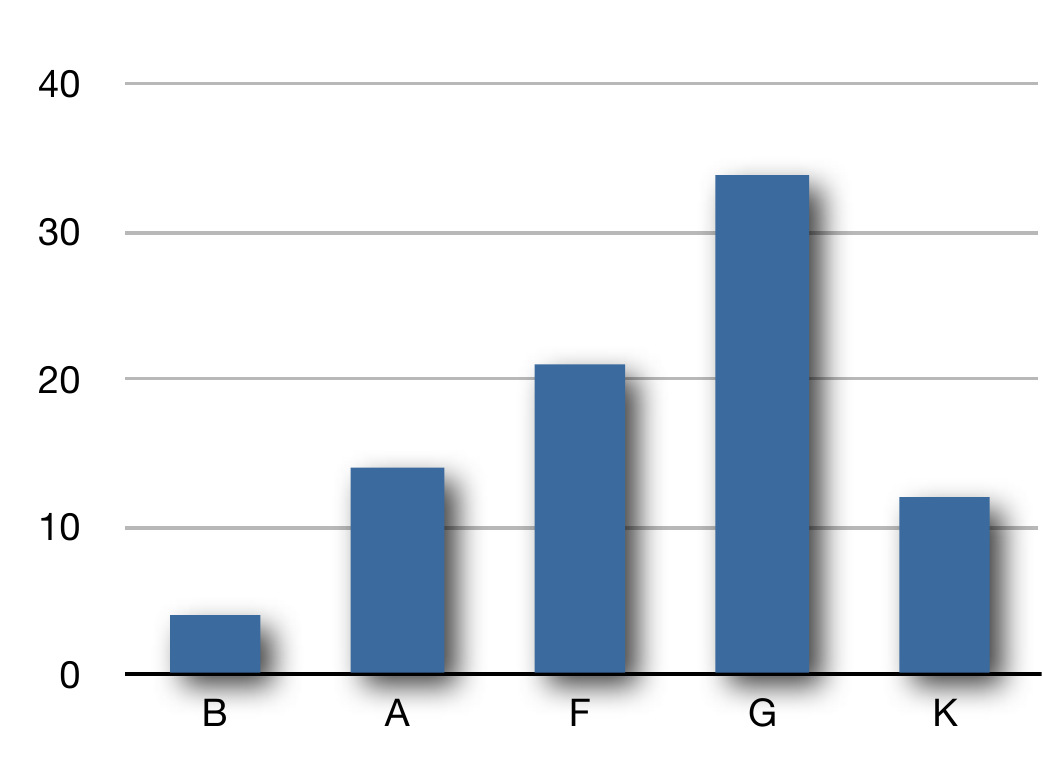}}
 \caption{Number of stars for each spectral types in the Lyot project survey initial sample. }
\label{SpectralTypes}
\end{figure}

In Table 1, we list all stars for which we obtained usable data (the spectral-type distribution of the initial sample is shown in Figure \ref{SpectralTypes}).  Some of these stars have known binaries or disks and we did not include binary targets in the non-detection analysis (therefore including 58 stars listed in Table \ref{dr_result_table}  with the detection threshold achieved).  Binary stars in this survey will be separately studied in a future paper.

The observing procedure involved first acquiring the target star into the AO loop and optimizing the AO system parameters (AO bandwidth, tip-tilt rates, atmospheric dispersion correction).  Next,  several unocculted ``core'' images of the target star, i.e. with the star $\sim$1000-1500 mas away from the occulting mask, were obtained in $H$ band.  The core images were used for accurate photometry and the calculation of coronagraphic dynamic range as discussed later in this work.  Once a set of non-saturated ``core'' images was successfully obtained, the star was aligned with the coronagraphic mask using several large ($\sim$100 mas) offsets applied to the FSM. After the star was occulted by the mask, our internal tip-tilt  system kept the star aligned in place. Finally, several long (30s-120s) exposures were obtained in all six polarimetric modes. The typical total exposure time for a target is 20 minutes. Note that the first observing runs did not include the polarimetric mode.

\begin{deluxetable}{c||ccc|cccccc}

\tabletypesize{\footnotesize}

\startdata
Star Name& Spec Type  & Dist & H mag & Date& Exp Time &Field Rot &Base  & $R_{paralactic}$   & $R_{\mathrm{Coude}}$   \\
&  &  (pc) &  & & (s) & ($^\circ$) &Exp Time (s)&  ('')  & ('')  \\
HIP7345 & A1V & 61.3 & 5.53 & 2006.12.12 &  1020. & 14.1 &  60. & 0.49 & 2.17\\
HIP8102 & G8V & 3.7 & 1.80 & 2006.12.12 &  1860. & 14.9 &  60. & 0.46 & 1.91\\
 &  &  &  & 2006.12.20 &  2160. & 9.9 &  360. & 0.69 & 1.64\\
HIP16537 & K2V & 3.2 & 1.88 & 2004.09.20 &  2122. & 39.7 &  20. & 0.02 & 1.89\\
 &  &  &  & 2004.09.22 &  560. & 6.2 &  10. & 1.10 & 6.01\\
 &  &  &  & 2005.01.24 &  1845. & 26.7 &  20. & 0.26 & 1.08\\
 &  &  &  & 2005.01.27 &  630. & 5.7 &  20. & 1.21 & 1.28\\
 &  &  &  & 2006.12.20 &  4320. & 39.8 &  240. & 0.17 & 0.53\\
HIP19859 & G0V & 21.2 & 5.01 & 2006.12.19 &  1440. & 12.1 &  120. & 0.57 & 1.21\\
HIP22910 & A0V & 144.0 & 5.06 & 2005.01.24 &  2400. & 44.9 &  240. & 0.15 & 0.73\\
 &  &  &  & 2006.02.13 &  1440. & 36.2 &  120. & 0.19 & 0.55\\
 &  &  &  & 2006.12.12 &  3960. & 53.0 &  60. & 0.13 & 0.29\\
 &  &  &  & 2006.12.19 &  3600. & 32.4 &  300. & 0.21 & 1.71\\
HIP23143 & A3 & 131.0 & 6.26 & 2006.12.19 &  720. & 14.7 &  120. & 0.47 & 22.02\\
HIP25813 & B5V & 88.5 & 4.60 & 2005.01.24 &  431. & 18.9 &  8. & 0.36 & 0.88\\
HIP27288 & A2IV & 21.5 & 3.31 & 2006.12.19 &  360. & 2.6 &  60. & 2.70 & 226.84\\
HIP27913$^\mathrm{a}$ & G0V & 8.7 & 3.19 & 2005.01.26 &  200. & 0.5 &  50. & 78.13 & 5.25\\
HIP29800 & F5IV-V & 19.6 & 4.20 & 2005.01.27 &  1180. & 17.4 &  120. & 0.40 & 1.35\\
HIP32362 & F5IV & 17.5 & 1.81 & 2006.12.12 &  1080. & 19.5 &  60. & 0.35 & 1.10\\
 &  &  &  & 2006.12.19 &  1440. & 13.8 &  240. & 0.50 & 2.29\\
HIP33212 & F8V & 31.0 & 5.52 & 2005.01.24 &  1932. & 51.7 &  120. & 0.13 & 0.49\\
HIP34116 & B0IV & 247.0 & 6.22 & 2006.12.19 &  2880. & 18.1 &  120. & 0.38 & 0.85\\
HIP36366 & F0V & 18.5 & 3.16 & 2005.01.26 &  1120. & 20.6 &  20. & 0.33 & 0.69\\
 &  &  &  & 2006.12.12 &  360. & 6.9 &  60. & 0.99 & 5.82\\
HIP37826 & K0III & 10.3 & -0.85 & 2006.12.14 &  75. & 2.2 &  10. & 0.06 & 0.31\\
HIP38228 & G5IV & 21.8 & 5.36 & 2005.01.27 &  1448. & 45.5 &  120. & 0.15 & 1.06\\
 &  &  &  & 2006.12.12 &  1440. & 37.3 &  120. & 0.18 & 0.81\\
HIP43587 & G8V & 12.5 & 4.26 & 2005.01.26 &  1320. & 27.3 &  60. & 0.25 & 0.81\\
HIP44127 & A7V & 14.6 & 2.76 & 2005.01.27 &  750. & 26.0 &  30. & 0.26 & 0.71\\
HIP44248 & F5V & 16.4 & 3.08 & 2006.12.16 &  540. & 7.0 &  20. & 0.99 & 2.78\\
HIP44897 & F9V & 19.1 & 4.60 & 2006.12.14 &  1200. & 14.2 &  200. & 0.49 & 1.68\\
HIP45170 & G9V & 20.5 & 4.77 & 2006.12.12 &  1320. & 56.8 &  120. & 0.02 & 2.00\\
HIP46509 & F6V & 17.1 & 3.58 & 2005.01.27 &  1230. & 9.9 &  30. & 0.69 & 6.08\\
HIP46843 & K1V & 17.7 & 5.24 & 2006.12.19 &  720. & 16.4 &  120. & 0.42 & 2.59\\
HIP47080 & G8III & 11.2 & 3.72 & 2006.12.16 &  1440. & 19.9 &  120. & 0.35 & 1.33\\
 &  &  &  & 2006.12.19 &  720. & 10.2 &  120. & 0.68 & 5.81\\
HIP48113 & G0.5V & 18.4 & 3.73 & 2006.12.19 &  720. & 6.1 &  120. & 1.14 & 3.39\\
HIP49081 & G3V & 14.9 & 4.04 & 2006.12.19 &  720. & 10.4 &  120. & 0.66 & 2.71\\
HIP49669 & B7V & 23.8 & 1.66 & 2005.01.26 &  1560. & 54.8 &  10. & 0.13 & 0.41\\
HIP50564$^\mathrm{a}$ & F6IV & 21.2 & 3.94 & 2006.12.16 &  720. & 22.8 &  120. & 0.30 & 2.04\\
HIP53721 & G1V & 14.1 & 3.74 & 2006.12.16 &  1440. & 20.9 &  120. & 0.33 & 1.70\\
HIP54745 & G0V & 21.7 & 5.02 & 2005.01.26 &  1140. & 18.6 &  60. & 0.37 & 1.01\\
HIP56809 & G0V & 23.3 & 5.11 & 2005.05.18 &  1320. & 16.5 &  60. & 0.42 & 1.40\\
HIP57632 & A3V & 11.1 & 1.92 & 2005.01.22 &  2374. & 104.8 &  25. & 0.02 & 0.71\\
HIP57757 & F9V & 10.9 & 2.36 & 2005.05.17 &  1550. & 17.9 &  50. & 0.38 & 1.29\\
 &  &  &  & 2005.05.18 &  1710. & 20.3 &  45. & 0.34 & 0.95\\
HIP60074 & G2V & 28.5 & 5.61 & 2006.12.12 &  2940. & 3.1 &  60. & 2.22 & 0.73\\
 &  &  &  & 2006.12.14 &  2600. & 4.3 &  200. & 1.60 & 0.89\\
HIP61174 & F2V & 18.2 & 3.37 & 2006.12.19 &  1800. & 11.0 &  150. & 0.63 & 1.61\\
HIP61498 & A0V & 67.1 & 5.79 & 2005.01.26 &  1320. & 9.2 &  120. & 0.02 & 4.57\\
HIP62523 & G5V & 17.2 & 4.71 & 2007.06.08 &  1080. & 41.5 &  90. & 0.17 & 2.24\\
HIP62956 & A0 & 24.8 & 1.70 & 2007.06.13 &  720. & 27.7 &  60. & 0.25 & 0.42\\
HIP64792 & G0V & 18.0 & 4.11 & 2007.06.13 &  720. & 2.5 &  120. & 2.78 & 2.14\\
HIP65721 & G5V & 18.1 & 3.46 & 2005.05.18 &  780. & 28.8 &  60. & 0.24 & 24.54\\
HIP66249 & A3V & 22.4 & 3.15 & 2004.06.12 &  480. & 5.6 &  30. & 1.22 & 5.78\\
 &  &  &  & 2005.05.17 &  810. & 12.8 &  60. & 0.54 & 3.37\\
 &  &  &  & 2007.06.06 &  360. & 2.9 &  60. & 2.39 & 6.16\\
 &  &  &  & 2007.06.12 &  1080. & 18.6 &  60. & 0.37 & 0.61\\
HIP67927 & G0IV & 11.3 & 1.53 & 2005.05.18 &  1310. & 333.5 &  10. & 0.02 & 1.40\\
HIP70873 & G5V & 23.6 & 4.85 & 2005.05.18 &  1440. & 17.0 &  60. & 0.41 & 1.13\\
HIP71284 & F2V & 15.5 & 3.46 & 2007.06.13 &  720. & 10.3 &  120. & 0.67 & 2.40\\
HIP72659$^\mathrm{a}$ & G8V & 6.7 & 2.25 & 2005.05.17 &  1261. & 320.8 &  30. & 0.02 & 0.54\\
HIP72848$^\mathrm{a}$ & K2V & 11.5 & 4.17 & 2007.06.13 &  720. & 2.8 &  120. & 2.43 & 1.91\\
HIP73996 & F5V & 19.7 & 4.01 & 2005.05.18 &  1200. & 9.2 &  60. & 0.86 & 0.76\\
HIP74975 & F8IV & 24.7 & 3.95 & 2007.06.13 &  720. & 6.3 &  120. & 1.10 & 3.62\\
HIP77257 & G0V & 11.8 & 3.07 & 2007.06.08 &  720. & 11.0 &  60. & 0.63 & 2.64\\
HIP77542 & B9.5 & 99.0 & 6.86 & 2007.06.06 &  1440. & 20.5 &  120. & 0.34 & 1.45\\
 &  &  &  & 2007.06.08 &  2880. & 26.4 &  120. & 0.26 & 1.19\\
 &  &  &  & 2007.06.12 &  720. & 5.7 &  120. & 1.20 & 5.20\\
HIP77622 & A2 & 21.6 & 3.44 & 2005.05.14 &  1685. & 15.8 &  60. & 0.44 & 0.68\\
HIP77760 & F8V & 15.9 & 2.74 & 2007.06.13 &  720. & 6.8 &  120. & 1.02 & 3.20\\
HIP78072 & F6IV & 11.1 & 2.88 & 2007.06.13 &  720. & 5.9 &  120. & 1.17 & 2.08\\
HIP78459 & G0V & 17.4 & 3.99 & 2005.05.17 &  1620. & 41.0 &  60. & 0.17 & 1.29\\
HIP79248 & K0V & 18.1 & 4.80 & 2007.06.08 &  1440. & 14.6 &  120. & 0.47 & 1.18\\
HIP79672 & G2V & 14.0 & 4.16 & 2007.06.13 &  360. & 2.8 &  60. & 2.50 & 10.36\\
HIP81300 & K0V & 9.8 & 4.05 & 2007.06.12 &  720. & 6.7 &  120. & 1.03 & 8.54\\
HIP84862 & G0V & 14.4 & 3.90 & 2007.06.12 &  720. & 12.6 &  120. & 0.55 & 3.88\\
HIP85042 & G5IV & 19.5 & 4.80 & 2007.06.12 &  720. & 6.1 &  120. & 1.13 & 5.12\\
HIP85653 & G5 & 22.3 & 5.47 & 2007.06.12 &  720. & 5.8 &  120. & 1.19 & 3.41\\
HIP86032 & A5III & 14.3 & 1.72 & 2004.06.12 &  170. & 3.1 &  17. & 2.17 & 29.33\\
 &  &  &  & 2007.06.12 &  240. & 2.5 &  10. & 2.76 & 1.14\\
HIP86400 & K3V & 10.7 & 4.40 & 2005.05.14 &  1700. & 24.2 &  100. & 0.29 & 1.33\\
HIP86974 & G5IV & 8.4 & 1.56 & 2005.05.17 &  900. & 29.5 &  10. & 0.23 & 2.86\\
HIP87819 & A1V & 122.0 & 5.53 & 2007.06.08 &  1440. & 8.4 &  120. & 0.82 & 6.41\\
HIP88601 & K0V & 5.1 & 1.88 & 2005.05.17 &  780. & 15.2 &  15. & 0.45 & 39.41\\
HIP88745 & F7V & 15.7 & 3.24 & 2007.06.12 &  120. & 2.4 &  10. & 2.93 & 8.62\\
HIP88972 & K2V & 11.1 & 4.46 & 2007.06.12 &  720. & 6.7 &  120. & 1.03 & 2.57\\
HIP89474 & G2V & 22.7 & 4.84 & 2005.05.15 &  1380. & 15.1 &  60. & 0.46 & 1.13\\
HIP91262 & A0V & 7.8 & -0.03 & 2004.06.11 &  347. & 18.2 &  0.5 & 0.38 & 1.76\\
 &  &  &  & 2005.05.14 &  2245. & 43.5 &  8. & 0.16 & 0.53\\
 &  &  &  & 2007.06.06 &  240. & 6.1 &  10. & 1.12 & 5.60\\
HIP93017 & F9V & 15.0 & 3.61 & 2005.05.15 &  280. & 10.7 &  20. & 0.83 & 104.34\\
HIP93966 & G4V & 21.0 & 4.56 & 2007.06.12 &  720. & 30.5 &  120. & 0.23 & 3.19\\
HIP94076 & G1V & 49.0 & 5.29 & 2004.06.13 &  960. & 14.1 &  120. & 0.49 & 6.11\\
 &  &  &  & 2007.06.08 &  360. & 5.9 &  60. & 1.15 & 6.79\\
HIP95319 & G8V & 15.5 & 4.74 & 2007.06.12 &  720. & 14.2 &  120. & 0.49 & 3.08\\
HIP95447 & G8IV & 15.1 & 3.33 & 2007.06.12 &  720. & 5.8 &  120. & 1.19 & 2.28\\
HIP96183$^\mathrm{a}$ & G5V & 20.2 & 5.25 & 2007.06.12 &  720. & 1.4 &  120. & 4.82 & 1.87\\
HIP96441 & F4V & 18.6 & 3.72 & 2004.06.11 &  79. & 1.7 &  5. & 4.03 & 14.99\\
 &  &  &  & 2005.05.14 &  1440. & 16.4 &  60. & 0.42 & 0.96\\
HIP97295 & F7V & 20.9 & 3.98 & 2007.06.12 &  720. & 14.0 &  120. & 0.49 & 3.23\\
HIP97649 & A7V & 5.1 & 0.10 & 2007.06.08 &  300. & 13.2 &  10. & 0.52 & 2.60\\
HIP98767 & G7IV-V & 15.9 & 4.24 & 2005.05.15 &  1680. & 48.5 &  60. & 0.14 & 1.13\\
 &  &  &  & 2007.06.06 &  2160. & 30.7 &  120. & 0.22 & 1.73\\
HIP98819 & G0V & 17.7 & 4.43 & 2005.05.15 &  1140. & 23.5 &  30. & 0.29 & 2.12\\
HIP99031 & K0IV & 24.2 & 3.45 & 2007.06.12 &  720. & 7.5 &  120. & 0.92 & 2.56\\
HIP100970$^\mathrm{a}$ & G3IV & 37.4 & 5.32 & 2007.06.08 &  720. & 0.8 &  120. & 8.32 & 1.89\\
HIP102488 & K0III & 22.1 & 0.10 & 2007.06.12 &  600. & 13.0 &  20. & 0.53 & 2.43\\
HIP104214 & K5V & 3.5 & 2.54 & 2007.06.08 &  1440. & 10.2 &  120. & 0.68 & 3.45\\
HIP114570 & F0V & 24.5 & 3.76 & 2006.12.12 &  720. & 48.3 &  60. & 0.14 & 0.47\\
HIP116771 & F7V & 13.8 & 2.99 & 2006.12.19 &  720. & 8.3 &  120. & 0.83 & 13.09

\enddata

\tablecomments{$^a$ These targets have declinations within 2 degrees of the latitude of
the observatory. Therefore a point source in the field of view (FOV) rotates significantly only close to the zenith and can be subtracted. As it does not affect a significant number of stars, this was not taken into account in the reduction routine but each raw image was checked by eye for these targets.}

\label{LogTable}

\end{deluxetable}

\subsection{Data Analysis and Angular Differential Imaging}

\subsubsection{Data Pipeline} \label{sec:pipeline}

\begin{figure}[htbp]
\center
\resizebox{0.5\hsize}{!}{\includegraphics{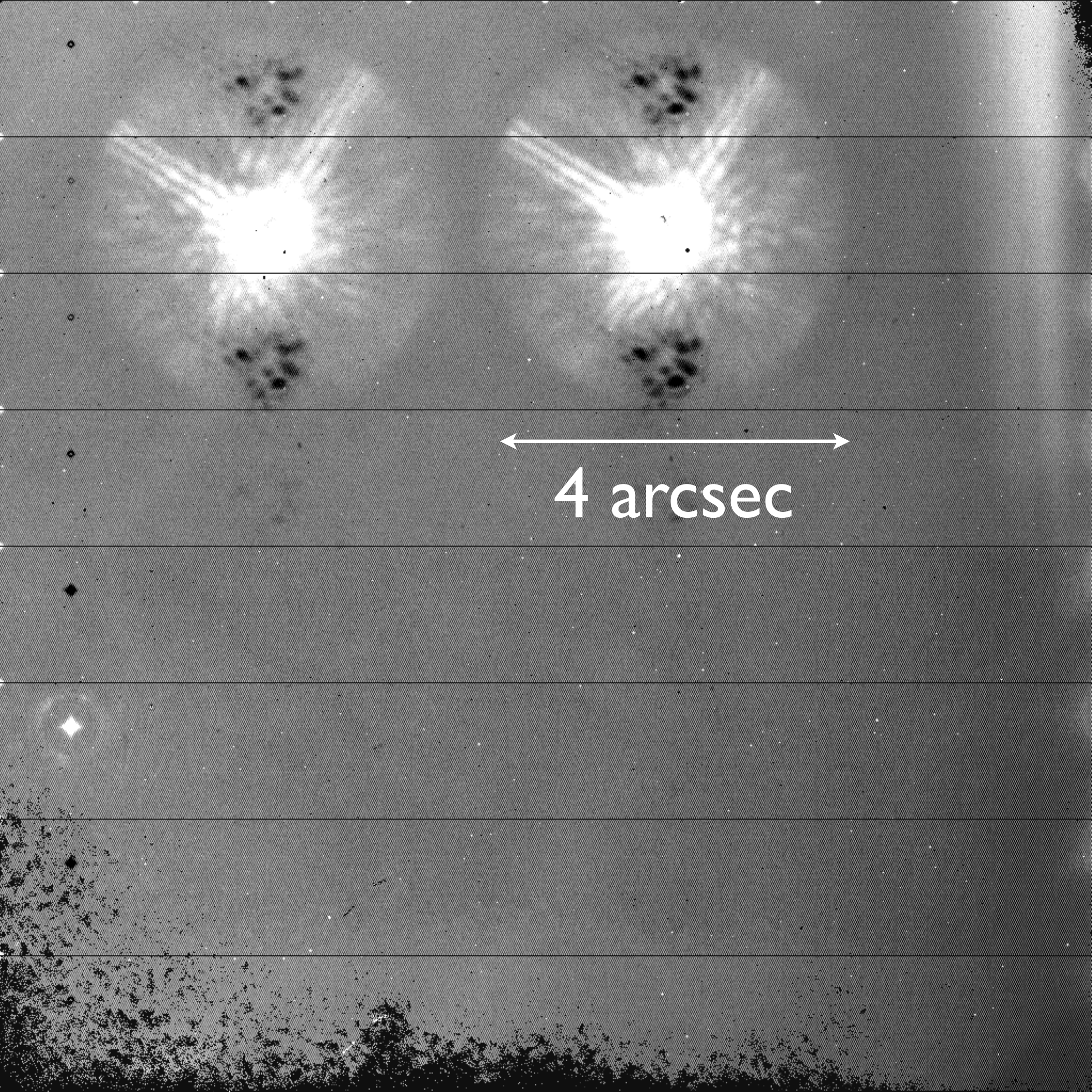}}
  \caption{Example of coronagraphic-polarimetric raw data with the Lyot project instrument. The two fields correspond to two orthogonal polarizations as separated by the Wollaston prism. A complete observation consists of a series of three such frames, with different polarizations. This shows one quadrant of the engineering-grade Hawaii-2 chip used in Kermit, demonstrating that while the chip had extensive regions of bad pixels, we were able to place our FOV in a clean region of the chip.}
  \label{fig1}
\end{figure}

\begin{figure}[htbp]
\center
\resizebox{0.3\hsize}{!}{\includegraphics{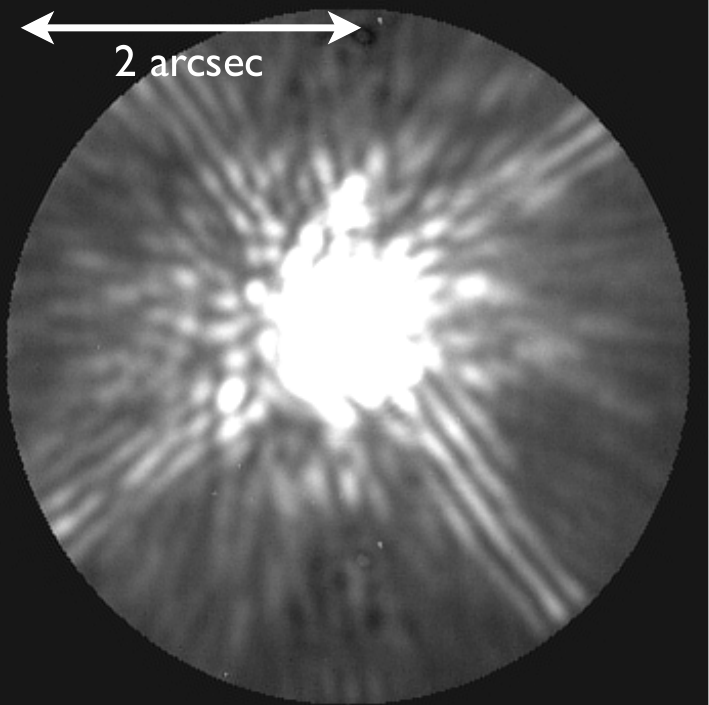}}
  \caption{Example of processed data for the unpolarized intensity (to be compared to Figure \ref{fig1}). Note that the Lyot project images are limited by a circular FOV, which is due to a circular field mask with a 2" OWA.}
  \label{fig4}
\end{figure}

The raw data reduction pipeline uses classical operations common to any astronomical data, and includes specific steps to correct effects from the detector or to process the polarimetric signal.  An example of raw data is given in Figure \ref{fig1} and shows several artifacts which must be corrected (bad pixels on the detector, periodic negative replication of the image and a checkerboard pattern due to the readout electronics). The two images correspond to two orthogonal polarization states that have been separated by the Wollaston prism.  In the case of companion searches, the polarimetric information is not used. Polarimetric imaging with this instrument is detailed by \citet{OBH08} and \citet{HOS09}.  An example of the reduced unpolarized image data is given in Figure \ref{fig4}, where all detector effects have been corrected.

\subsubsection{Angular Differential Imaging} \label{sec:adi}

\begin{figure*}[htbp]
\center
\resizebox{.8\hsize}{!}{\includegraphics{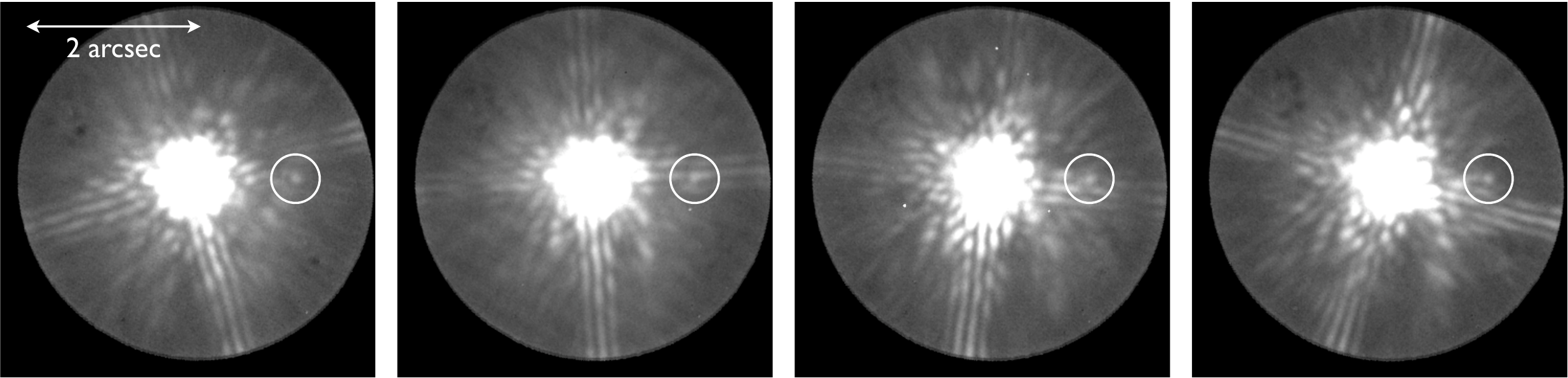}}
 \caption{Example of differential rotation between a point source (around HIP98767) at rest in the FOV and the diffraction pattern of the spiders in the telescope. This also illustrates how speckle noise can easily hide a point source. }
 \label{rotation}
\end{figure*}

In order to maximize image contrast, we used the Angular Differential Imaging (ADI) technique proposed by \citet{MLD06}. During the observation of a star the sky rotates with respect to the detector for an altitude/azimuth telescope, and this rotation is used for speckle noise suppression. In the case of the Lyot project, the camera is located at the Coud\'e focus and an additional rotation exists between the detector and the telescope pupil. We use this differential rotation between the objects in the FOV and the quasi-static speckles (illustrated by Figure \ref{rotation}) to isolate and subtract the speckles from the images. 
In order to use ADI, we need to compute the different rotation angles for each image. These rotation angles can readily be computed using the coordinate system transformation:
\begin{equation}\label{anglepap}
\cos(\theta_{\mathrm{para}})=\frac{\sin(\phi)-\sin(\mathrm{Alt})sin(\delta)}{\cos(\mathrm{Alt})\cos(\delta)},
\end{equation}
where $\phi$ denotes the latitude of the telescope, $\delta$ the declination of the target and $\mathrm{Alt}$ the altitude of the target in the sky at the time of the observation. Also $\theta_{\mathrm{para}}$ has the same sign as the hour angle (HA), so it can be unambiguously determined from its cosine.
In our case (the signs are highly dependent on the number of reflections in the instrument) the rotation of the field at the Coud\'e focus is given by
\begin{equation}\label{angle}
\theta_{\mathrm{Coude}}=\theta_{\mathrm{para}}-\mathrm{Alt}+\mathrm{Az}+\theta_{0},
\end{equation}
where $\mathrm{Alt}-\mathrm{Az}$ corresponds to the second rotation caused by leaving the frame of the telescope and $\theta_{0}$ is an offset angle due to the orientation of the camera. This offset angle as well as the correct combination of signs has been calibrated by observing well-known astrometric binaries and finding the combination of signs that aligns the binaries with the right position angle.

For each reduced and registered image of an ADI sequence, a reference coronagraphic point-spread function (PSF, diffraction pattern of an unresolved on-axis source -coronagraphic PSF-, which includes the quasi-static speckles due to imperfections of the optics), must be built from the images of this sequence. The way in which this reference coronagraphic PSF is built directly affects the degree of noise attenuation possible. To construct the reference coronagraphic PSF we simply take the median of all the images of the sequence after alignment in the frame of the telescope (which means that the diffraction pattern due to the spiders holding the secondary mirror is stationary during the whole sequence).

If enough field rotation has occurred during the sequence, a point source has moved by at least twice its FWHM and will be largely rejected by the median. Thus, only the average coronagraphic stellar PSF is left. The minimum radial separation at which this occurs is noted as $R_{\mathrm{para}}$ and is given by
\begin{equation}\label{rayon}
R_{\mathrm{para}}=\frac{2 \cdot \mathrm{FWHM}}{\theta_{\mathrm{para}}(t=t_{\mathrm{tot}})-\theta_{\mathrm{para}}(t=0)},
\end{equation}
where $t_{\mathrm{tot}}$ is the duration of the whole observation. In our study we chose $\mathrm{FWHM}=\lambda/D=121\,\mathrm{mas}\approx 8.9$ pixels.

Since the median is taken over a large number of images, the pixel-to-pixel noise (i.e., PSF, flat field, dark and sky Poisson noises, and detector readout noise) of the reference image is much less than that of any individual image. Thus, this method minimizes the speckle noise and should give a residual image where sensitivity is limited by pixel-to-pixel noise, but relies strongly on the assumption that quasi-static speckles are correlated during the whole sequence.
The coronagraphic PSF is subtracted from each image of the sequence, which gives a new sequence of images without most of the quasi-static speckles. As shown by \citet{HOS07} these speckles are produced by optical imperfections in the telescope and instrument, and that their typical lifetime can be up to an hour. ADI can be refined and optimized for long exposures \citep{LMD07}. However, in the Lyot project case where observations typically last 20-30 minutes with two differential rotations and additional centering issues due to the use of a coronagraph, we use the basic ADI method.

Because the Lyot project coronagraph is at the Coud\'e focus of an altitude/azimuth telescope, the speckles created after the telescope rotate at a different rate. This includes speckles created by the coronagraph optics and the AO system dead actuators \citep{OSP05}.  Therefore, we reapply a second ADI subtraction in the CCD camera frame, which is at rest with respect to the coronagraph. 
Because we used two successive ADI subtractions associated with different rotation angles, we introduce a second minimum radial separation at which a potential point source would not be subtracted ($R_{\mathrm{Coude}}$). The order in which the two ADI subtractions are performed is set by $R_{\mathrm{para}}\leq R_{\mathrm{Coude}}$, 
so that the first subtraction does not remove potential point sources in the domain of validity of the second subtraction. In some cases where $R_{\mathrm{Coude}}$ can be larger than the FOV, only one subtraction is performed. 

\begin{figure*}[htbp]
\center
\resizebox{.6\hsize}{!}{\includegraphics{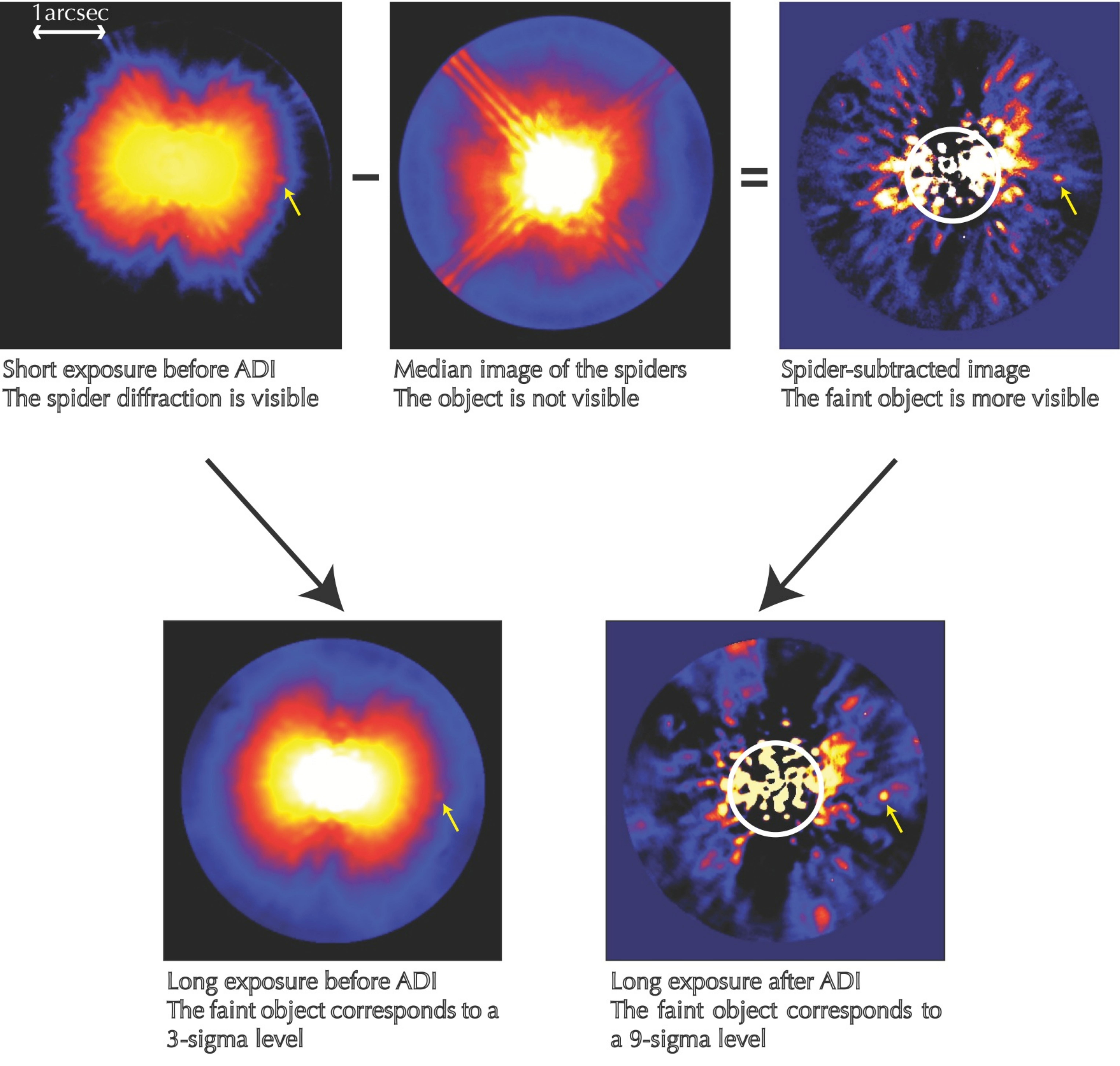}}
 \caption{Illustration of ADI processing with only one rotation: a reference median image of the PSF is obtained from the sequence of short exposure images (the field point source is rejected by the median). Then, this image is subtracted to each image of the sequence and the residuals are added to average the uncorrelated noise. Here the images are shown with different stretch to illustrate the principle.}
 \label{adi}
\end{figure*}

After ADI subtractions, we add up all the images of the sequence to average the residual noise and improve the contrast of a potential companion. As we do that with the three sequences, we end up with three averaged images (summarized in Figure \ref{adi}):
\begin{itemize}
	\item Unprocessed averaged image.
	\item Averaged image after the first subtraction. Valid for separations bigger than $R_{\mathrm{para}}$.
	\item Averaged image after the second subtraction. Valid for separations bigger than $R_{\mathrm{Coude}}$.
\end{itemize}

\subsubsection{Issues: Centering of the star}\label{sec:centering}

Measuring the position of the star in coronagraphic images is a complex problem, mainly due to the fact that the star is hidden behind the mask of the coronagraph, and the coronagraphic PSF is not a simple Airy pattern. Therefore, the position of the star cannot be simply deduced from the peak or the shape of the PSF \citep{DHO06}, as can be done for standard ADI.
\citet{DHO06} also suggest that in the case of the Lyot project, the tip-tilt sensor information can be used for centering information. However this requires a reference image where both the position of the star and of the tip-tilt sensor are known. This is possible for well- known binaries, for which the position of the occulted star can be deduced from the position of its companion, but since this reference has to be known each time the telescope moves, this technique is not practical for the observation of a single star. Thus, we achieved the centering of the star using the coronographic PSF symmetry as discussed by \citet{DHO06} who found it to be accurate at the 37 mas level.
\citet{SO06,MLM06} propose practical solutions to overcome this astrometric problem in coronagraphic images. Next generation high contrast instruments such as the Gemini Planet Imager will include these techniques \citep{MGP08}.
We tested the approach proposed by \citet{SO06} involving a periodic grid of wires with known width and spacing in a pupil plane ahead of the occulting coronographic FPM to produce fiducial images of the obscured star at known locations relative to the star. 
This device allows the observer to find the center and obtain photometry of the occulted star.  However the prototype tested for the Lyot project coronograph suffered from strong image distortion occurring at the edge of the FOV (where the fiducial images are located) and the information from the fiducial images was not directly usable. More recent tests of the astrometric grid on the Gemini Planet Imager testbed fully demonstrate the technique (A. Sivaramakrishnan et al., private communication 2008).
In the case of the Lyot data, we simply assumed that the star is aligned with the center of the occulting mask. 

\section{Dynamic Range and Performance Analysis}

\subsection{Sensitivity estimation}\label{sec:sens-est}

In order to characterize the completeness of the survey we evaluate the detection limits of the observations. The dynamic range (DR) corresponds to the faintest companion that can be detected at a given position in the field, at the detection limit. The procedure to estimate the noise in a real image is not unique and often contains several free parameters. There are also many choices for signal, as well as the S/N level to set. 

The DR is a measurement of the total flux ratio between the primary star and the faintest detectable companion at an angular separation $r$. The natural signal to consider is the total flux of a point source at this position and the noise has to be computed consistently on a PSF scale (approximately $\lambda /D$). Such methods possess free parameters (e.g. size of the boxes to consider) that will change the value of the DR.
\begin{figure}[htbp]
\center
\resizebox{0.8\hsize}{!}{\includegraphics{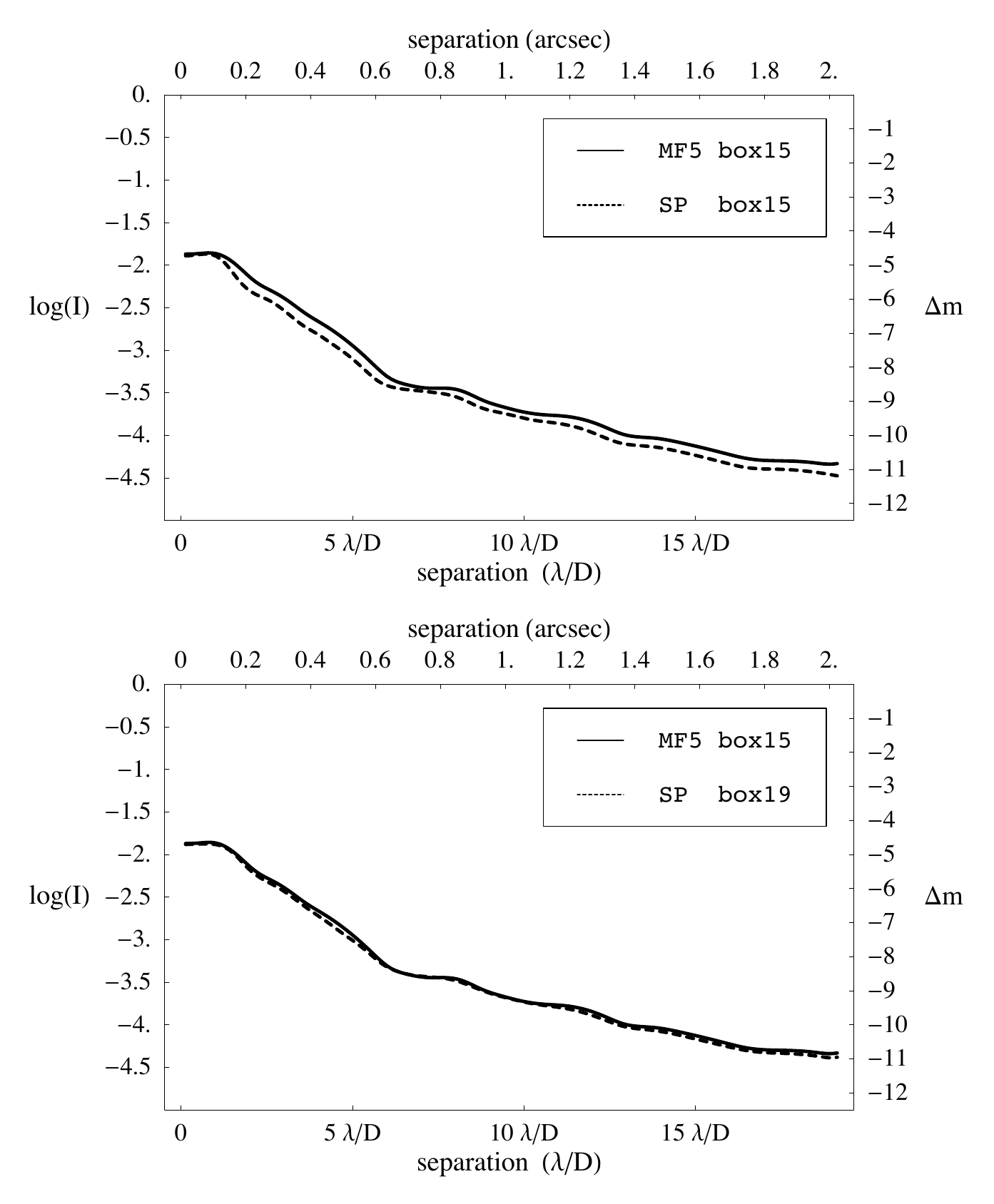}}
  \caption{Comparison of the results with the pixelwise method (dashed curve) and the matched filter (solid curve) for the same box size (top) and for different box sizes below. The matched filter curve shows a total agreement with the pixelwise method for box sizes slightly smaller. The S/N Ratio (here fixed to 5) level can also be chosen to rescale and customize the noise estimation routine to our data (see Section  \ref{SNRlevel}).}
  \label{fig12}
\end{figure}
We tested several DR measurement methods (in particular, a pixelwise measurement and a Matched Filter method that consists of computing the correlation between the image and a PSF and to compute the noise in the resulting image. Thus it takes the shape of a point source into account in the computation of the PSF-scale noise map and pixelwise noise created by e.g. "hot-pixels" is removed. See \citet{SOH06} for a detailed definition.), but each one has free parameters that can affect the results by a factor 1.5 to 2. Recently, \citet{HHK08} used a PSF fitting technique which is an improved Matched Filter that can take into account a slow background trend. Nevertheless, it appears from our tests that the \emph{shape} of the radial profile is quite independent of the method (See Figure \ref{fig12}). Therefore, we decided to choose the following method (which is not computationally intensive) and to normalize it afterward by choosing a relevant S/N ratio (see Section \ref{SNRlevel}).

We use a pixelwise dynamic range where the reference signal is defined as the maximum value of the unocculted stellar PSF \textrm{(with scaled exposure times)}.  We assume here that the companion is not affected by the coronagraph, which is the case with a classical Lyot coronagraph, not too close to the coronagraphic mask \citep{LGK01,LS05,SOH06}. Therefore the dynamic range results are valid beyond the inner working angle (IWA), comparable to the mask size. The IWA is approximately 0.3arcsec for the Lyot project data. 
The noise has to be estimated from the image itself, and in the case where the speckle noise dominates, it is necessary to measure this noise locally.  The spatial scale of a speckle is similar to the scale of the PSF and therefore it is necessary to measure the noise over a region that contains several realizations of speckles.  

This can be done in several ways. In order to minimize the effect of non-radial pattern such as the diffraction by the spiders or the ghosts created by the astrometric grid, we choose a two-dimensional measurement estimating the noise locally in a box centered at each position in the field. This box should be small enough to account for the local variations of the dynamic range, but large enough to include several realizations of speckles. For these data the sampling is 8.9 pixel per $\lambda/D$ ($\lambda/D=121$ mas and the pixel scale is 13.51 mas per pixel). Then we had to consider box sizes of a few $\lambda/D$ on a side and chose a 21 pixel wide box. We take the median of this "noise map" along circles of increasing radii to derive a radial DR profile. The median rejects extreme values, so geometrical patterns such as spider diffraction do not lead to an overestimation of the noise in the whole image.
As discussed above, ADI processing is not valid closer than a separation of $R_{\mathrm{para}}$ for the first rotation and $R_{\mathrm{Coude}}$ for the second one. In order to have one DR valid for the whole image, we computed the DR for our three output images (see Figure \ref{DR}) and combined them into a composite DR with three different zones delimited by $R_{\mathrm{para}}$ and $R_{\mathrm{Coude}}$ (see Figure \ref{DR}).

\subsection{Signal-to-noise ratio determination}\label{SNRlevel}

The statistical significance of a given signal depends not only on the standard deviation of the background noise, but also on the statistical distribution which best describes that noise.
For a Gaussian distribution of an uncorrelated noise, the confidence interval (CI) corresponding to a 99.9 \% confidence level (CL) has a width of 5 times the standard deviation ($\sigma$) of the distribution. For this reason, a S/N level of 5 is commonly chosen. But, for the same CL, the width of the CI for an exponential distribution is 9$\sigma$.

\begin{figure*}[htbp]
\center
\resizebox{.8\hsize}{!}{\includegraphics{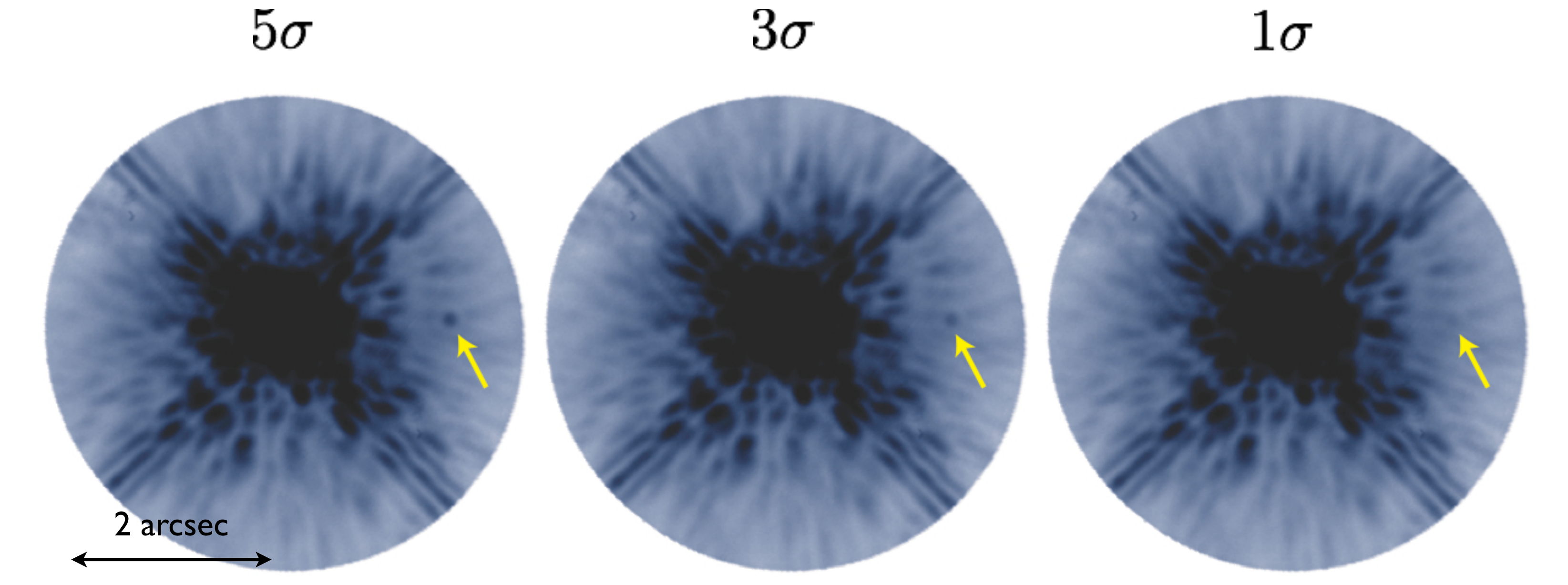}}
 \caption{Injection of fake companions at different signal-to-noise ratio to determine a detection threshold. A 3$\sigma$ point source is considered detected.}
 \label{eyeball}
\end{figure*}

In our images the noise distribution is not Gaussian and the noise is correlated from one pixel to another \citep{AS04,FG06,HOS07,SFA07,MLM08}. In addition, the signal has a shape (Airy pattern) and can be more easily detected by the eye (this is what a matched filter tries to reproduce). As shown in Figure \ref{eyeball}, a 5$\sigma$ point source is clearly visible for a trained eye. Therefore, the S/N level should take into account the specific properties of our data, the detection capacities of the eye, and rescale the noise estimation routine that we have chosen. We wish to stress here that since the distribution of the noise is different from one instrument and reduction pipeline to another, the chosen S/N level does not have an absolute meaning and cannot be compared directly from one survey to another.

The S/N level for our study was chosen by estimating the noise in several images with our routine (as explained in Section \ref {sec:sens-est}) and by injecting fake sources with different S/N. A blind test was carried out afterward among the seven members of the data reduction team to decide the S/N corresponding to a visible point source with less than 5$\%$ false positives. This process converged around an S/N level of 3$\sigma$ (central image of Figure \ref{eyeball}). Taking a security margin, we used a 3.5$\sigma$ detection level. This level is a conservative limit as sources as low as 1$\sigma$ were detected during the test and the data searching (see below). Since this determination has been conducted with our noise estimation routine on our data, this S/N level accounts for all the particularities of our routine, and normalizes the dynamic range to the real sensitivity limit of our instrument. 

We stress here that the 3.5$\sigma$ detection level used here is not our only detection criterion. It is merely an alarm threshold. Once a spot in the final image with an S/N$>3.5$ is detected, its position and shape are checked in each initial images. As opposed to speckles, point sources have a circular PSF, do not fluctuate or disappear, and are at rest in the sky frame during the whole sequence. 
All the images were searched by eye. Over the entire survey, only one point source in the substellar luminosity regime was detected around HIP98767 with:S/N $\approx 1$ in each raw image (upper left panel of Figure \ref{adi}; the point source is already visible), S/N $\approx 3$ in the averaged image and S/N $\approx 9$ in the final processed one (lower right panel). This is the only source that was detected to be brighter than the detection threshold in a final composite image \textbf{and} rotating with the plane of the sky during the whole observing time. This shows that our detection criterion (S/N $>3.5$) in addition to the tracking of the point source during a single observation yields a low false positive probability at most $<1\%$. This candidate observed at a later epoch has been confirmed to be a background source as it was no longer visible.  As mentioned earlier, we did not include binary stars in our non-detection analysis.

\subsection{Results}\label{sec:results}

\begin{figure}[htbp]
\center
\resizebox{.6\hsize}{!}{\includegraphics{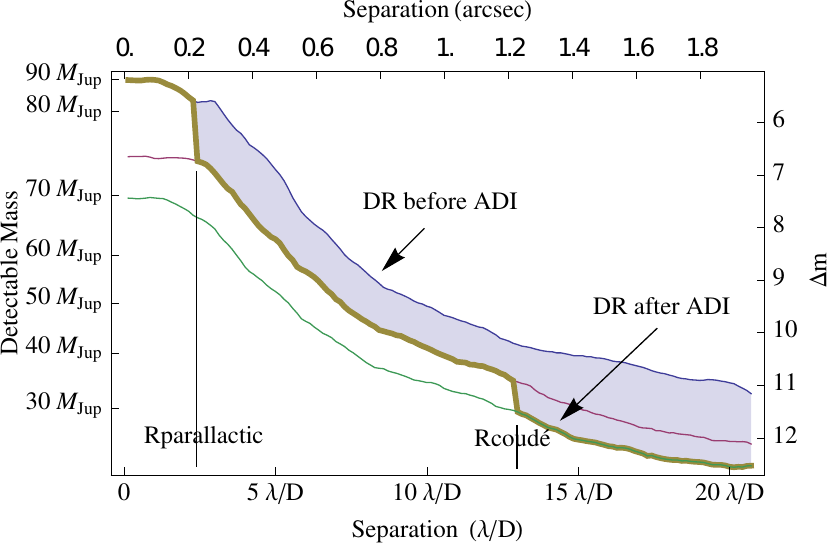}}
 \caption{Composite dynamic range (sensitivity limit) obtained with a $3.5 \sigma$ S/N ratio in $\Delta$magnitude in the $H$ band (brown thick line) around HIP91262. Three dynamic range profiles are calculated before and after two successive ADI subtractions corresponding to the two rotation angles at the Coud\'e focus. The three dynamic range curves are combined according to their validity zones to produce the composite detection curve. The detectable masses have been computed with the mass-luminosity relation (see Section \ref{CBmodel}) given by \citet{BCA03} for a Solar type star of 1.3 Gyr at 11 pc. Here, the ADI allows the detection of a 30$\mjup$ companion at 1."4 which would not have been detected in the unprocessed image.}
 \label{DR}
\end{figure}

The results of our sensitivity estimation in the $H$ band are given for all the stars of the survey in Table \ref{dr_result_table}. Figure \ref{DR} shows the case of HIP91262 where enough observing time was devoted to the star for the ADI to be working properly . The average improvement of the dynamic range yielded by the ADI noise reduction routine is about 1-2 mag, and the best contrast achieved in the $H$ band is about $\Delta H=$13.5 (which corresponds to a contrast of $10^{-5.4}$) at 2". Over the whole survey, average contrasts achieved span between 9$\lesssim \Delta H\lesssim$11 at 1$"$ and between  11$\lesssim \Delta H\lesssim$13 at 2" (with our $3.5\sigma$ detection level). This allows us to detect a companion down to 30$\mjup$ around the median target of our sample (a G8 star located 20pc away), and down to 15$\mjup$ around the most favorable cases.

\begin{deluxetable}{c|cccccccccccccc}
\tabletypesize{\footnotesize}

\tablehead{\colhead{Star Name}  \vline&\multicolumn{9}{c}{Dynamic Range ($\Delta$ mag in $H$ band)} \vline &Spec&Distance&$H$ Mag&Age$^a$\\ 
\colhead{}  \vline&\multicolumn{9}{c}{Separation}  \vline&Type & (pc) & & (Gyr) \\
\colhead{} \vline & \colhead{0.4''} & \colhead{0.6''} & \colhead{0.8''} & \colhead{1''} & \colhead{1.2''} & \colhead{1.4''} & \colhead{1.6''} & \colhead{1.8''} & \colhead{2.0''}   \vline&  & & & } 

\startdata
HIP19859 & 6.24 & 7.28 & 7.73 & 8.51 & 11.3 & 11.5 & 11.7 & 11.5 & 11.4 & G0V & 21.2 & 5.01 & 0.9 \\
HIP27288 & 6.54 & 7.45 & 8.22 & 8.95 & 9.4 & 9.78 & 10.1 & 10.3 & 10.4 & A2IV & 21.5 & 3.31 & 0.4 \\
HIP29800 & 6.48 & 7.6 & 8.89 & 10.5 & 11. & 11.4 & 11.7 & 11.9 & 11.9 & F5IV-V & 19.6 & 4.2 & 0.6 \\
HIP32362 & 6.84 & 7.95 & 11.1 & 11.5 & 12. & 12.3 & 12.7 & 12.8 & 13.1 & F5IV & 17.5 & 1.81 & 1.6 \\
HIP36366 & 7.25 & 8.78 & 10.3 & 10.7 & 11.2 & 11.8 & 12.6 & 12.9 & 13.1 & F0V & 18.5 & 3.16 & 0.7 \\
HIP37826 & 6.38 & 7.69 & 9.04 & 9.42 & 9.98 & 10.6 & 10.8 & 10.8 & 11. & K0III & 10.3 & -0.845 & 1.4 \\
HIP44248 & 6.98 & 7.69 & 8.53 & 9.14 & 9.54 & 9.99 & 10.4 & 10.5 & 10.7 & F5V & 16.4 & 3.08 & - \\
HIP44897 & 5.94 & 7.58 & 8.76 & 9.36 & 11.3 & 11.5 & 11.5 & 11.5 & 11.7 & F9V & 19.1 & 4.6 & 2.2 \\
HIP46509 & 6.54 & 7.54 & 9.95 & 10.7 & 11.3 & 11.3 & 11.7 & 12.1 & 12.2 & F6V & 17.1 & 3.58 & 0.3Ê \\
HIP47080 & 6.59 & 7.14 & 11.2 & 11.8 & 12.2 & 12.4 & 12.3 & 12.3 & 12.3 & G8III & 11.2 & 3.72 & 1.2 \\
HIP48113 & 6.19 & 7.61 & 9.04 & 9.58 & 10.1 & 10.5 & 11. & 11.2 & 11.2 & G0.5V & 18.4 & 3.73 & 5.1 \\
HIP49081 & 5.96 & 7.34 & 8.65 & 9.41 & 9.94 & 11.9 & 11.9 & 11.6 & 11.6 & G3V & 14.9 & 4.04 & 6.9 \\
HIP50564 & 6.51 & 7.13 & 11. & 11.3 & 11.8 & 12. & 12. & 11.8 & 12. & F6IV & 21.2 & 3.94 & 1.5 \\
HIP53721 & 6.98 & 7.81 & 11.2 & 11.5 & 11.9 & 12.3 & 12.1 & 12. & 12.1 & G1V & 14.1 & 3.74 & 7.3 \\
HIP54745 & 6.66 & 8.32 & 10.1 & 10.4 & 10.9 & 11.5 & 11.8 & 12.1 & 12.3 & G0V & 21.7 & 5.02 & 0.5 \\
HIP56809 & 6.01 & 7.55 & 8.44 & 9.93 & 10.5 & 10.9 & 11.2 & 11.3 & 11.5 & G0V & 23.3 & 5.11 & 0.1 \\
HIP61174 & 6.33 & 7.72 & 9. & 9.67 & 10.3 & 12.6 & 12.5 & 12.3 & 12.2 & F2V & 18.2 & 3.37 & 0.5 \\
HIP62523 & 8.06 & 9.12 & 9.87 & 10.6 & 11. & 11.1 & 11.3 & 11.5 & 11.5 & G5V & 17.2 & 4.71 & 7.2 \\
HIP62956 & 8.04 & 10.4 & 11.5 & 12.6 & 12.9 & 13.2 & 13.4 & 13.4 & 13.3 & A0 & 24.8 & 1.7 & 0.5 \\
HIP64792 & 6.32 & 7.39 & 8.53 & 9.18 & 9.91 & 10.1 & 10.3 & 10.8 & 11.4 & G0V & 18. & 4.11 & 5.4 \\
HIP65721 & 6.3 & 8.78 & 9.51 & 10.1 & 10.6 & 11. & 11.1 & 11.4 & 11.6 & G5V & 18.1 & 3.46 & 7.5 \\
HIP66249 & 5.59 & 6.58 & 7.46 & 7.96 & 10.3 & 10.8 & 11.2 & 11.5 & 11.7 & A3V & 22.4 & 3.15 & 0.6 \\
HIP67927 & 7.24 & 8.59 & 9.18 & 10. & 10.4 & 10.9 & 11.4 & 11.6 & 11.9 & G0IV & 11.3 & 1.53 & 1.8 \\
HIP70873 & 6.12 & 7.75 & 8.73 & 10.2 & 10.6 & 11.1 & 11.2 & 11.4 & 11.7 & G5V & 23.6 & 4.85 & 9.1 \\
HIP71284 & 7.61 & 8.97 & 9.85 & 10.5 & 11.1 & 12.6 & 13.1 & 13.2 & 13.3 & F2V & 15.5 & 3.46 & 1.2 \\
HIP72659 & 7.87 & 9.22 & 9.86 & 10.6 & 11.5 & 12.1 & 12.7 & 13. & 13.3 & G8V & 6.7 & 2.25 & 4.5 \\
HIP73996 & 6.47 & 7.42 & 8.17 & 9.31 & 9.63 & 10.3 & 11.8 & 12.1 & 12.4 & F5V & 19.7 & 4.01 & 1.9 \\
HIP74975 & 7.16 & 8.16 & 9.3 & 9.9 & 10.4 & 10.6 & 10.9 & 11.2 & 12. & F8IV & 24.7 & 3.95 & 3.0 \\
HIP77257 & 5.87 & 7.37 & 8.72 & 9.35 & 10.3 & 11.2 & 11.6 & 11.7 & 12. & G0V & 11.8 & 3.07 & 6.6 \\
HIP77622 & 6.36 & 7.61 & 8.85 & 9.53 & 10.9 & 11.2 & 11.5 & 12.1 & 12.4 & A2 & 21.6 & 3.44 & 0.5 \\
HIP77760 & 6.88 & 7.67 & 8.7 & 9.32 & 9.76 & 10. & 10.3 & 10.7 & 11.3 & F8V & 15.9 & 2.74 & 8.4 \\
HIP78072 & 6.5 & 7.32 & 8.24 & 8.92 & 9.44 & 9.57 & 9.76 & 10.2 & 10.8 & F6IV & 11.1 & 2.88 & 3.6 \\
HIP78459 & 7.22 & 8.35 & 9.54 & 10. & 10.6 & 11. & 11.1 & 11.3 & 11.7 & G0V & 17.4 & 3.99 & 10 \\
HIP79672 & 7.11 & 8.21 & 9.35 & 9.93 & 10.4 & 10.7 & 11. & 11.3 & 11.9 & G2V & 14. & 4.16 & 6.9 \\
HIP81300 & 6.04 & 7.28 & 8.72 & 9.43 & 9.97 & 10.5 & 10.7 & 11. & 11.3 & K0V & 9.78 & 4.05 & 8.2åÊ \\
HIP84862 & 6.94 & 7.97 & 9.03 & 9.64 & 11. & 11.6 & 11.6 & 11.7 & 11.9 & G0V & 14.4 & 3.9 & 5.8 \\
HIP85042 & 6.11 & 7.53 & 8.85 & 9.1 & 9.95 & 10.6 & 10.7 & 11.1 & 10.7 & G5IV & 19.5 & 4.8 & 9.4 \\
HIP86032 & 5.78 & 7.17 & 8.71 & 9.33 & 10.1 & 10.7 & 11. & 11.2 & 11.4 & A5III & 14.3 & 1.72 & 0.7 \\
HIP86400 & 5.72 & 7.36 & 9.15 & 9.68 & 10.2 & 10.7 & 11. & 11.2 & 11.6 & K3V & 10.7 & 4.4 & 2.4 \\
HIP88601 & 6.06 & 7.27 & 8.15 & 10.4 & 10.8 & 11. & 11.1 & 11.4 & 11.7 & K0V & 5.09 & 1.88 & 1.3 \\
HIP88745 & 7.21 & 8.7 & 9.05 & 9.8 & 10.4 & 10.4 & 10.4 & 10.6 & 10.6 & F7V & 15.7 & 3.24 & 11 \\
HIP88972 & 5.92 & 7.46 & 8.67 & 9.19 & 10.2 & 10.6 & 10.8 & 11.2 & 11. & K2V & 11.1 & 4.46 & 1.5 \\
HIP89474 & 6. & 6.66 & 7.47 & 10.6 & 10.9 & 11.4 & 11.5 & 11.8 & 12. & G2V & 22.7 & 4.84 & 8.6 \\
HIP93966 & 6.56 & 9.18 & 9.86 & 10.5 & 11. & 11.3 & 11.4 & 11.6 & 11.7 & G4V & 21. & 4.56 & 9.9 \\
HIP94076 & 7.1 & 8.23 & 9.29 & 10.1 & 10.5 & 10.9 & 11. & 11.1 & 10.9 & G1V & 49. & 5.29 & - \\
HIP95319 & 6.66 & 7.81 & 8.68 & 9.27 & 11. & 11.2 & 11.3 & 11.3 & 11.4 & G8V & 15.5 & 4.74 & 7.0 \\
HIP95447 & 7.14 & 8.17 & 9.19 & 9.92 & 10.7 & 11.3 & 11.4 & 11.8 & 11.8 & G8IV & 15.1 & 3.33 & 10.3 \\
HIP96183 & 6.13 & 7.58 & 8.32 & 9.02 & 9.67 & 10.4 & 10.6 & 10.9 & 10.3 & G5V & 20.2 & 5.25 & 7.3 \\
HIP96441 & 6.38 & 7.46 & 8.52 & 10. & 10.5 & 10.8 & 11.2 & 11.5 & 12.3 & F4V & 18.6 & 3.72 & 1.0 \\
HIP97295 & 7.1 & 8.19 & 9.24 & 9.79 & 11.2 & 11.4 & 11.6 & 11.7 & 11.9 & F7V & 20.9 & 3.98 & 2.4 \\
HIP97649 & 6.5 & 7.49 & 9.13 & 9.59 & 11.2 & 11.6 & 11.9 & 12.1 & 12.3 & A7V & 5.14 & 0.102 & 0.6 \\
HIP98767 & 7.59 & 8.62 & 9.49 & 10. & 10.4 & 10.9 & 11.1 & 11.4 & 11.8 & G7IV-V & 15.9 & 4.24 & 11 \\
HIP99031 & 6.58 & 8.18 & 9.14 & 9.67 & 10.6 & 10.9 & 11. & 11.2 & 12.1 & K0IV & 24.2 & 3.45 & 5.2 \\
HIP100970 & 6.92 & 8.13 & 9.33 & 10.1 & 10.8 & 11.4 & 11.7 & 11.9 & 11.4 & G3IV & 37.4 & 5.32 & 8.4 \\
HIP102488 & 7.21 & 8.33 & 9.14 & 9.85 & 11.3 & 11.6 & 11.8 & 11.9 & 12.2 & K0III & 22.1 & 0.104 & 2.7 \\
HIP104214 & 8.05 & 9.04 & 10.8 & 11.5 & 12.1 & 12.7 & 12.7 & 12.8 & 13. & K5V & 3.48 & 2.54 & 0.5 \\
HIP114570 & 6.93 & 8.14 & 9.1 & 9.59 & 10.3 & 10.6 & 11.1 & 11.1 & 11.3 & F0V & 24.5 & 3.76 & 0.9 \\
HIP116771 & 6.6 & 7.69 & 8.43 & 9.14 & 9.76 & 10.1 & 10.4 & 11.9 & 11.8 & F7V & 13.8 & 2.99 & 3.5 \\
\enddata

\tablecomments{$^a$ When age could not be evaluated, a value of 10 Gyr was used in the statistical analysis.}
\label{dr_result_table}

\end{deluxetable}


\section{Survey statistical analysis}\label{survey}

\subsection{Statistical Formalism}\label{Bayes}

In order to place constraints on the properties of exoplanets from null results we convert the information on the sensitivity into more physical parameters and incorporate them into a statistical study. 
Following the Bayesian approach of \citet{LDM07}, we consider the observation of $N$ stars enumerated by $\mathrm{j}=1,...,N$. Let $n_{\mathrm{i}}(I)$ be the fraction of stars having $\mathrm{i}$ companions of mass and SMA in the interval $I=\left[m_\mathrm{min},m_\mathrm{max} \right]\bigcap\left[a_\mathrm{min},a_\mathrm{max}\right]$.
The completeness $p_\mathrm{j}(I)$ is the probability that a companion around the star $\mathrm{j}$ would be detected given the detection limits of the observations (The computation of this quantity is discussed in Section \ref{sec:completeness}).
The probability of not detecting any companion around star $\mathrm{j}$ ($P_{\mathrm{nd,j}}$) is given by
\begin{equation}\label{non detect}
P_{\mathrm{nd,j}}(I)=\sum^{\infty}_{\mathrm{i}=0} n_\mathrm{i}(I)\cdot (1-p_\mathrm{j}(I))^{\mathrm{i}}.
\end{equation}
This formal definition is not directly usable. 

Precedent authors avoided this infinite sum by not considering multiple planetary systems. The recent direct observation of a multiple system by \citet{MMB08} shows that including the possibility of planetary systems in our analysis is of great importance.
The point of this section is to \textit{demonstrate} that allowing for the presence of multiple planets, we can rigorously use standard formalism of \citet{LDM07} and Equation (\ref{CI3}). 

 We introduce the fraction of stars that have \textit{at least} one companion ($f(I)=\sum^{\infty}_{\mathrm{i}=1} n_{\mathrm{i}}(I)$) in the interval of mass and SMA considered. Let us also introduce $n_\mathrm{i}^*=n_\mathrm{i}/f$ (for $\mathrm{i}\geqslant1$). Thus:
$
n_{0}(I)=1-f(I)
$
and
$
\sum^{\infty}_{\mathrm{i}=1} n_{\mathrm{i}}^*(I)=1.
$
The knowledge of the fraction of stars $n_\mathrm{i}^*$ having exactly $i$ companions  in the interval $I$ is not directly accessible. However this set of quantities is not required to constrain the population of extrasolar planets, which will be described with the parameter $f$. It is sufficient to know that these $n_i^*$ exist and are $>0$. With these notations, $p_\mathrm{nd,j}$ becomes:
\begin{eqnarray}
P_{\mathrm{nd,j}} &=&\sum^{\infty}_{\mathrm{i}=0} n_{\mathrm{i}}\cdot (1-p_\mathrm{j})^{\mathrm{i}}  \nonumber\\
&=&(1-f)+f(1-p_\mathrm{j})\sum^{\infty}_{\mathrm{i}=1}  n_\mathrm{i}^*\cdot (1-p_\mathrm{j})^{\mathrm{i}-1} \nonumber\\
&=&(1-f)+f(1-p_\mathrm{j})\sum^{\infty}_{\mathrm{i}=1}  n_\mathrm{i}^*+f(1-p_\mathrm{j})\sum^{\infty}_{\mathrm{i}=1}n_\mathrm{i}^*\cdot [(1-p_\mathrm{j})^{\mathrm{i}-1}-1],
\end{eqnarray}
and since $\sum^{\infty}_{\mathrm{i}=1} n_{\mathrm{i}}^*=1$
\begin{eqnarray}\label{non detect3}
P_{\mathrm{nd,j}}&=&1-f\cdot p_\mathrm{j} + f(1-p_\mathrm{j})\sum^{\infty}_{\mathrm{i}=1}n_i^*\cdot [(1-p_\mathrm{j})^{\mathrm{i}-1}-1] \nonumber\\
&=&1-f\cdot p_\mathrm{j}-f\cdot p_\mathrm{j}^* 
\end{eqnarray}
 The problem of the definition of such infinite sums can be reasonably avoided by considering that there must be an upper limit for the number of planets having stable orbits around a single star.
It is easy to see that $p_\mathrm{j}^*$ is positive. \citet{LDM07} find that $P_{\mathrm{nd,j}}=1-f\cdot p_\mathrm{j}$ because they do not include the possibility of observing a multiple system. Indeed, $p_\mathrm{j}^*$ accounts for the fact that if several planets are present it is harder to \textbf{not} detect anything compared to a single-planet system. In other words, the probability of detecting \textit{at least} a companion increases with the number of planets in the system. 
Though the $p_\mathrm{j}^*$'s cannot be calculated without any assumption on the population of stars hosting a given number of planetary companions, we will demonstrate that ignoring this third term in Equation (\ref{non detect3}) yields perfectly rigorous bounds for $f$, which confirms the validity of the approach used in  \citet{LDM07}. 

Denoting $\left\{ d_\mathrm{j}\right\}$ the detections made by the observations, such that $d_\mathrm{j}$ equals 1 if at least one companion is detected around star $j$ or else equals 0, the likelihood of the data given $f$ is given by:
\begin{equation}\label{likelihood}
L(\left\{ d_\mathrm{j}\right\} | f)=\prod^{N}_{\mathrm{j}=1} (1-f(p_\mathrm{j}+p_\mathrm{j}^*))^{1-d_\mathrm{j}}\cdot(f(p_\mathrm{j}+p_\mathrm{j}^*))^{d_\mathrm{j}}.
\end{equation}
A proper determination of the probability that the fraction of stars having at least one companion is $f$ requires the use of Bayes' theorem:
\begin{equation}\label{densityprob}
p(f | \left\{ d_\mathrm{j} \right\})=\frac{L(\left\{ d_\mathrm{j} \right\} | f) \cdot p(f)}{\int^{1}_{0} L(\left\{ d_\mathrm{j} \right\} | f) \cdot p(f) df},
\end{equation}
where $p(f)$ is the \textit{a priori} probability density of $f$, called the prior distribution, and $p(f|\left\{ d_\mathrm{j} \right\})$ is the probability density of $f$ updated in light of the data, called the posterior distribution. Without any prior knowledge about $f$, we use the most ignorant prior distribution $p(f)=1$ (which means that, without prior knowledge, every value of the fraction $f$ is equiprobable). Given a CL $\alpha$, we can use the posterior distribution $p(f | \left\{ d_\mathrm{j} \right\})$ to determine a CI for $f$, bounded by $f_\mathrm{min}$ and $f_\mathrm{max}$  which simplifies in our case where there is no detection to $f_\mathrm{min}=0$ and
\begin{eqnarray}\label{CI2}
\alpha&=&\int^{f_\mathrm{max}}_0 p(f | \left\{ d_\mathrm{j} \right\}) df \nonumber \\
&=&\frac{\int^{f_\mathrm{max}}_0\prod^{N}_{\mathrm{j}=1} (1-f(p_\mathrm{j}+p_\mathrm{j}^*)) df}{\int^1_0\prod^{N}_{\mathrm{j}=1} (1-f(p_\mathrm{j}+p_\mathrm{j}^*)) df}
,
\end{eqnarray}
that can be seen as an implicit equation on $f_\mathrm{max}$. In other words, once an $\alpha$ is chosen and Equation (\ref{CI2}) is solved, we can say that the true fraction of stars that have at least one companion in our sample is less than $f_\mathrm{max}$ ($f\in\left[0,f_\mathrm{max}\right]$) with an $\alpha\%$ confidence. As highlighted above, this integral cannot be calculated without any assumption on the distribution of the number of companions. Since our goal is to give an upper limit to the fraction of stars hosting companions,  we demonstrate in the Appendix that disregarding the $p_\mathrm{j}^*$'s to calculate integral \ref{CI2} yields a higher value for $\tilde{f}_\mathrm{max}$ such that $\left[0,\tilde{f}_\mathrm{max}\right]$ is a perfectly rigorous bracketing of the fraction $f$ of stars having \textit{at least} a companion with an $\alpha\%$ confidence ($\Big[0,f_\mathrm{max}\Big]\subseteq\left[0,\tilde{f}_\mathrm{max}\right]$).

Intuitively, this result comes from the fact that ignoring the $p_\mathrm{j}^*$'s, which is equivalent to neglecting the possibility of multiple systems, increases our non-detection probability (Equation (\ref{non detect3})) around any star. It acts in the same way as decreasing the sensitivity of the observations. Thus, using the CI given by
\begin{eqnarray}\label{CI3}
\alpha=\frac{\int^{\tilde{f}_\mathrm{max}}_0\prod^{N}_{\mathrm{j}=1} (1-fp_\mathrm{j}) df}{\int^1_0\prod^{N}_{\mathrm{j}=1} (1-fp_\mathrm{j}) df},
\end{eqnarray}
we are sure that the true fraction of stars that have at least one companion in this range $\left[m_\mathrm{min},m_\mathrm{max} \right]$ and $\left[a_\mathrm{min},a_\mathrm{max}\right]$ $f$ is lower than $\tilde{f}_\mathrm{max}$.
In this work, we chose a typical value of $\alpha=0.70$.

\subsection{Completeness calculation for each star of the survey}\label{sec:completeness}

The determination of the $p_\mathrm{j}$s (completeness of each observation) is a critical step in this analysis. The completeness - which is the fraction of companions that would have been detected in the interval $\left[m_\mathrm{min},m_\mathrm{max} \right]\bigcap\left[a_\mathrm{min},a_\mathrm{max}\right]$ if they have actually been there - is a practical translation of our dynamic range to physical parameters because its value depends on the detection limits of the observations, on the ages and distances of the systems and on the masses, SMAs and the orbital eccentricity distribution of the companions. In calculating the completeness it is also important to account properly for orbital inclination and phase, as these significantly affect the distribution of projected separations of an orbit of a given SMA.

\subsubsection{Mont\'e Carlo simulations}\label{MCsimu}

The completeness was calculated using a Monte Carlo approach (following \citet{Bro04}). In order to minimize the assumptions to be made on the distributions of masses and SMAs of the companions, 10$^4$ planets were generated by randomly sampling the orbital eccentricity,  the three astrometric angles for the orientation of the orbit (inclination, longitude of the ascending node, argument of the periastron) and the mean anomaly, for \textbf{each point} in the Semi Major Axis (SMA) \textit{vs} Mass space around each star. This provides us with a \textbf{completeness map} for each star of the survey.

For all of our calculations, the orbital eccentricity distribution was assumed to be flat with $0 \leq e \leq 0.8$ in agreement with the radial velocity exoplanets sample \citep{BWM06}. The inclination angle has a constant distribution in $\sin(i)$, while the longitude of the ascending node, the argument of the periastron and the mean anomaly are given by uniform distributions between 0 and $2\pi$.
The true anomaly is computed from the mean anomaly by solving kepler's  equation numerically, which gives us the position of the planet on its orbit.
The artificial companion is projected on the plane of sky and its angular separation from the star is evaluated using the distance of the primary.

\subsubsection{Mass-Luminosity Relation}\label{CBmodel}

The translation from the mass of our simulated planets into luminosity (in our case, magnitude in $H$ band) that we can compare to our detection limit to see whether the companion would be eventually detected relies on the evolutionary models from \citet{BCA03}, which provides a mass-luminosity relation for giant, non-irradiated planets. For our survey the orbital separation between the primary and the companion is large enough to neglect the contribution of the star to the thermal history of the planet and its flux in the IR.

\subsubsection{Completeness Maps for each stars}\label{completeresults}

\begin{figure*}[htbp]
 \centering
 \begin{minipage}[b]{0.32\linewidth}
 \centering
 \resizebox{1.\hsize}{!}{\includegraphics {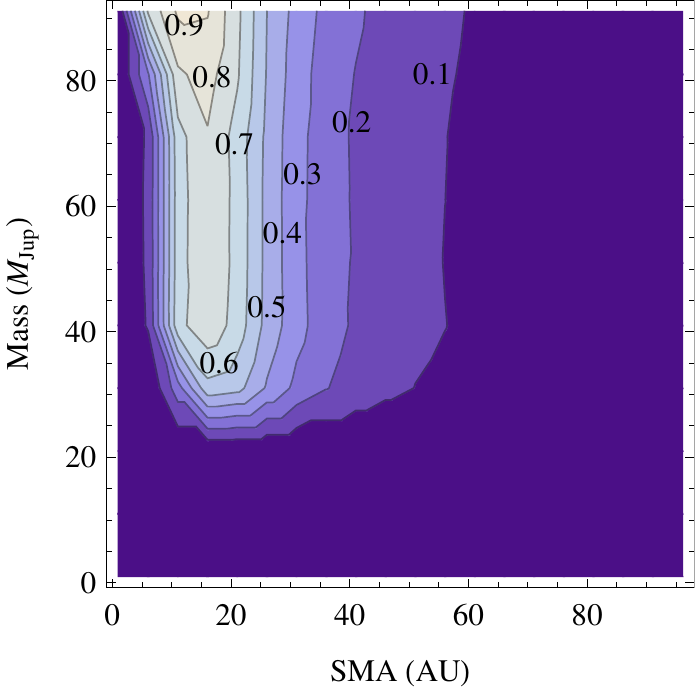}}
 \end{minipage}
 \begin{minipage}[b]{0.32\linewidth}
 \centering
 \resizebox{1.\hsize}{!}{\includegraphics {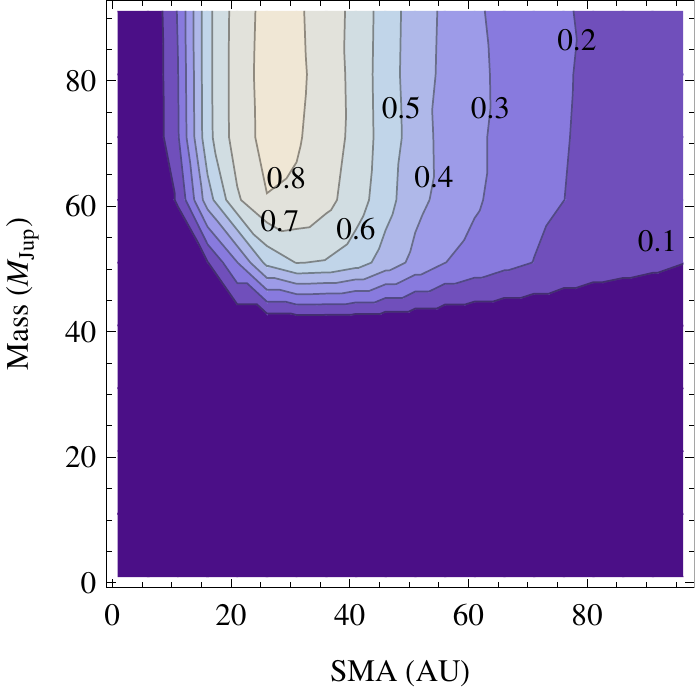}}
 \end{minipage}
 \begin{minipage}[b]{0.32\linewidth}
 \centering
 \resizebox{1.\hsize}{!}{\includegraphics {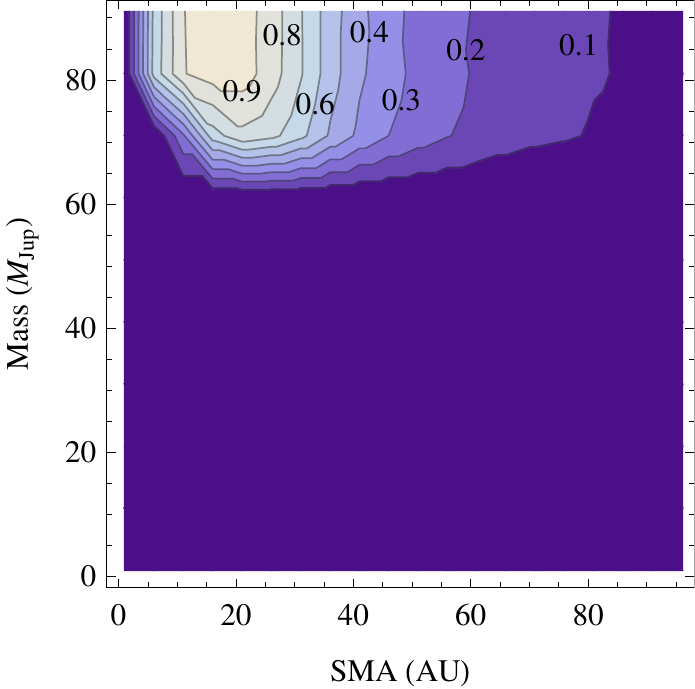}}
 \end{minipage}
 \caption{ Completeness of the observation of three stars from the survey in the Mass vs SMA space. From left to Right: HIP47080 (G8, 1.3 Gyr at 11 Parsec); HIP50564 (F6, 1.5 Gyr at 21 Parsec); HIP98767 (G7, 11 Gyr at 16 Parsec). For example, a contour region marker 0.7 means that 70 \% of the possible population of companion for this mass and SMA would have been detected}
 \label{mass vs sma completeness}
\end{figure*}

From our population of artificial companions we can identify those falling into the FOV of the coronagraph and bright enough to be detectable given our dynamic range profile. Computing the ratio of detectable companions to the total number of artificial companions for each point of the grid gives the completeness of the observation for this area of the parameter space. This provides us with \textbf{completeness maps} such as shown in Figure \ref{mass vs sma completeness}.

As illustrated by Figure \ref{mass vs sma completeness} these maps have similar shapes. There are two obvious patterns: 
\begin{itemize}
	\item At a given SMA, the completeness increases with the mass of the companion (bottom to top), which is explained by the fact that more massive planets are brighter and thus easier to detect.
	\item At a given mass, the completeness starts by increasing sharply with the SMA and decreases more slowly to zero at large SMAs. At short SMAs, this feature is due to the presence of the occulting mask, which completely hides the very close orbits. This creates a sharp cutoff of the completeness. At large SMAs less eccentric orbits are completely outside the FOV (that is the source of the decay). For the more eccentric orbits the companion can still be visible during a part of its orbit, even if Kepler's second law yields that the companion spends more time far from its star (hence the slower decay).
\end{itemize}
Even if these maps have common properties, Figure \ref{mass vs sma completeness} shows that the results vary from star to star depending on distance, age, and magnitude. Eventually, the completeness improves when the star is younger and closer because these factors tend to make the companions brighter. This is particularly visible with HIP98767 in Figure \ref{mass vs sma completeness} which is very old. But since the dynamic range is only a magnitude difference between the companion and the primary, the completeness also improves with faint stars. 

\section{Discussion}\label{sec:constraint}

\begin{figure}[htbp]
\center
\resizebox{.5\hsize}{!}{\includegraphics {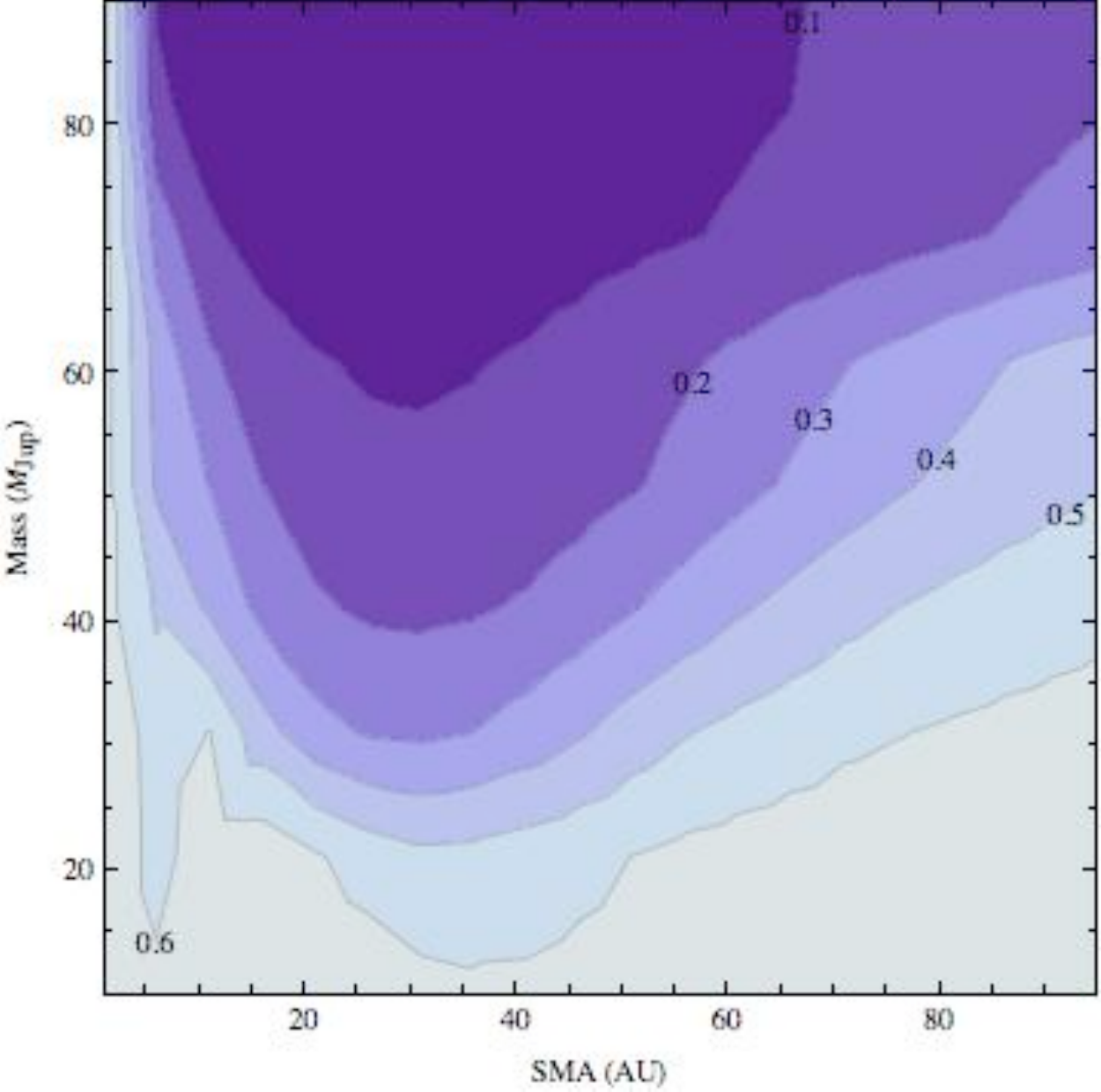}}
 \caption{Population constraint from the Lyot Survey in the Mass vs SMA space: Contours give the maximum companion frequency compatible with our observations, as a function of mass and SMA. For example, at SMA 30AU, no more than 30 \% of stars have a companion of 30$M_{\mathrm{Jup}}$ (with a 70\% CL). This figure looks like a negative image of the completeness maps. As expected, we place higher constraints on the population of companions where the sensitivity is high (i.e. for SMA between 10 and 60AU, and masses above 40MJup.}
\label{mass vs sma fmaxplot}
\end{figure}

Combining all these completeness maps giving the $p_\mathrm{j}$'s in the Bayesian approach discussed in Section \ref{Bayes} for each point of the parameter space, we obtain a map of the maximum companion frequency $f_\mathrm{max}$ for the stars of the survey (Figure \ref{mass vs sma fmaxplot}).
Figure \ref{mass vs sma fmaxplot} looks like a negative-image of the completeness maps of Figure \ref{mass vs sma completeness}. This is simply because constraints increase with completeness. Therefore, stronger constraints can be placed in the region of the parameter space where the observations are the most sensitive.
As expected, the survey is more sensitive to the more massive brown dwarfs and between 20 and 40 AUs which is consistent with the average distance of our targets ($\sim 20$ pc).

We can see from Figure \ref{mass vs sma fmaxplot} that the Lyot project survey is not very sensitive to companions under 20$\mjup$ (at a 70$\%$ significance level), which does not put strong constraints on giant Jupiter-like planets. This can partly be attributed to a bias in the selection of the targets. Because the AO system needs a lot of photons to work properly, our target list contains more bright stars (some A and even B spectral type stars) and fewer faint stars (K and M dwarfs) than the average population of nearby stars as shown in Figure \ref{SpectralTypes}. As discussed above, the sensitivity to small planets decreases when looking at bright stars. We derive an upper limit for the companion frequency of 30$\%$ more massive than 30$\mjup$ between 10 and 50 AU.

\begin{figure}[hbp]
\center
\resizebox{.5\hsize}{!}{\includegraphics {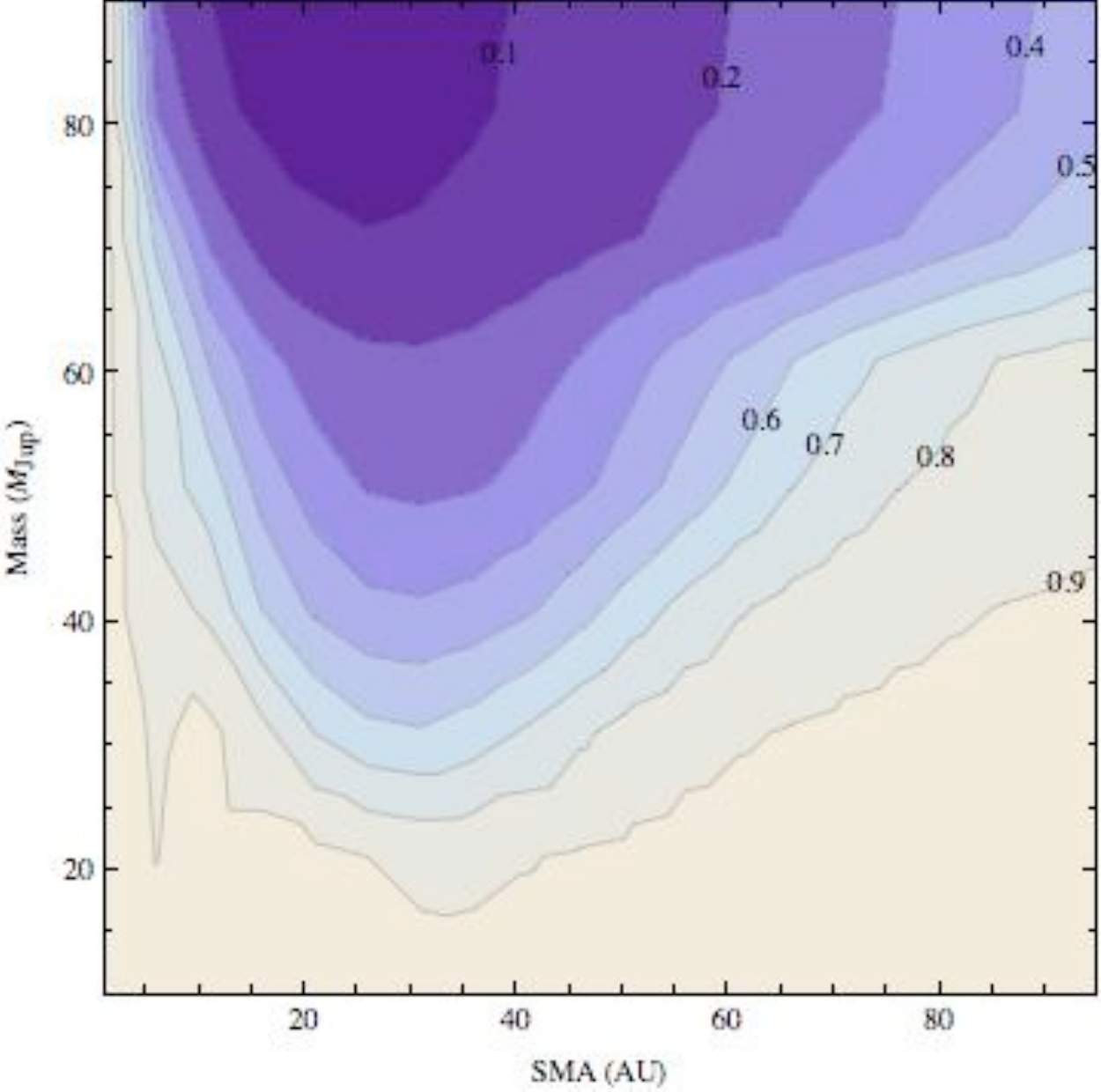}}
 \caption{Same as Figure \ref{mass vs sma fmaxplot} with a 95\% CL.}
\label{mass vs sma fmaxplot2}
\end{figure}

Some other studies concentrated on the careful analysis of null results of large-scale surveys such as this paper (\citet{NCB07,LDM07} and recently \citet{CLB09} and \citet{NC09}). Even without any detection of a planetary mass companion, these large surveys give useful information about the population of companions around nearby stars. Although these two studies do not use the same methodology, they yield similar conclusions.
\citet{NCB07} have compiled observations from \citet{MMH05} and \citet{BCM07} to form a sample of 60 nearby solar-like and low mass stars (the median star of the survey is a K2 at 25 pc). They stated that the fraction of stars with planets with SMA between 20 and 100 AU,
and mass above 4 $\mjup$, is 20\% or less with a 95\% confidence (for comparison, the maximum companion frequency is plotted with a 95\% significance level in Figure \ref{mass vs sma fmaxplot2}). For power-law mass distribution in mass and semi major axis ($\frac{\mathrm{d}N}{\mathrm{d}M} \propto M^{-1.16}$ in the range of 0.5-13 $\mjup$ and $\frac{\mathrm{d}N}{\mathrm{d}a} \propto a^{n}$) they also found an SMA upper cutoff of 18 AU for $n=0$, 48 AU for $n=0.5$, and 75 AU for $n=-0.61$ as proposed by \citet{CBM08}. \citet{LDM07} surveyed 85 low mass stars and are sensitive enough to detect planets more massive than 2 $\mjup$ with a projected separation in the
range 40-200 AU around their typical target: a 100 Myr old K0 star located 22 pc from the Sun. They considered similar power-law distributions ($\frac{\mathrm{d}N}{\mathrm{d}M} \propto M^{-1.2}$ and $\frac{\mathrm{d}N}{\mathrm{d}a} \propto a^{-1}$) and found that upper limits on the fraction of stars with at least one planet are 0.28 for $a$ in the range 10-25 AU, 0.13 for 25-50 AU, and 0.093 for 50-250 AU. Since our results do not place usable constraints in the range 0.5-13 $\mjup$ (mainly due to the fact that we observed bright massive stars due to the limitations imposed by the AO system), a comprehensive comparison with these studies is not needed, but our null results are in agreement with the conclusion of both papers. 

For large separations (20-2000 AU) \citet{MH09} and \citet{CMP09} constrained the population of substellar companions (see also \citealt{MH04,LBS05,CEB05,CES06,CHV09}). \citet{MH09} surveyed 266 F5-K5 stars and derived a companion frequency of $3.2^{+3.1}_{-2.7}\%$ in the 12-72 $\mjup$ mass range for separation between 28 and 1590 AU and $\frac{\mathrm{d}N}{\mathrm{d}M} \propto M^{-0.4}$. Because of our nondetection, we cannot constrain the lower limit for the companion frequency, but Figure \ref{mass vs sma fmaxplot} shows that our survey put weaker constrains and therefore confirms their results (in the 10-80 AU zone) that massive brown dwarfs are rare around solar-like stars in the range. In comparing the results of the Lyot project survey with other surveys, it is important to keep in mind that the telescope diameter is a relevant factor (3.6m for AEOS vs. 8m for Gemini or Very Large telescope). Although the extreme AO system on AEOS underperformed expectations, typical SR of $80\%$ in our $H$ band images helped compensate for a smaller diameter.

On the bright-star side, \citet{HHK08} obtained promising contrasts with deep AO images of Vega and $\epsilon$ Eri in the $L$\'\  (3.8
$\micron$) and $M$ (4.8 $\micron$) bands where better planet/star flux ratio are predicted, allowing the detection of exoplanets down to 6$\mjup$ around Vega. Unfortunately, the high sky background at these wavelengths seems to prevent these bands to be used for observation of fainter stars.

\section{Conclusion}
During the last few years, a tremendous amount of work has been devoted to direct imaging of extrasolar planets and brown dwarfs. This resulted on one hand in many successful detections for example by \citet{CLD04}, \citet{CLZ05}, \citet{NGW05}, \citet{BKC06}, \citet{SNS08}, \citet{KGC08}, \citet{MMB08} and \citet{LGC09}. Similar to this study, a few other large surveys attempting to constrain the population of substellar companions at large separations have been conducted \citep{NCB07,LDM07, CLB09}.  Our survey included 86 stars observed using extreme AO coronagraphic imaging. The survey included stars with spectral types ranging from B to K, and the median target is G8 at 20pc. The survey resolved the AB Aurigae disk \citep{OBH08}, the HR 4796A disk  \citep{HOS09}, and a number of binary stars, both newly discovered and previously known (to be presented in  future work).
 
The Lyot project observed and analyzed a sample of 58 star. The sample was not large enough to estimate the companion frequency for each stellar type and we considered the complete sample for the statistical analysis.  The constraints we are able to place based on this survey are limited to the brown dwarf regime at large separations. We confirm the rarity of massive brown dwarf around solar like and massive stars as we find that no more that 20\% of nearby stars have a companion with $M>40\mjup$ between 10 and 50 AU. 
Nevertheless, with the increasing number of surveys looking for exoplanets and in particular the upcoming Gemini Planet Imager and SPHERE, the constraints on the population of exoplanets are likely to grow stronger. It would then be interesting to compile all the detection and non-detection results in a unified statistical formalism to infer more comprehensive upper and lower limits to the population of exoplanets, as such knowledge is needed to constrain formation models.

The Lyot project was the first high-contrast coronagraph working with an extreme AO system. Its legacy extends beyond this survey results and other results \citep{OBH08,HOS09} as it helped develop a lot of techniques and expertise that have been very useful for other high contrast projects like Project 1640 \citep{HOB08} or Gemini Planet Imager \citep{MGP08}.

\acknowledgments
The Lyot project is based upon work supported by the
National Science Foundation under grant nos. 0334916,
0215793, and 0520822, as well as grant NNG05GJ86G
from the National Aeronautics and Space Administration
under the Terrestrial Planet Finder Foundation Science
Program. The Lyot project gratefully acknowledges
the support of the U.S. Air Force and NSF in creating
the special Advanced Technologies and Instrumentation
opportunity that provides access to the AEOS telescope.
Eighty percent of the funds for that program are provided
by the U.S. Air Force. This work is based on observations
made at the Maui Space Surveillance System,
operated by Detachment 15 of the U.S. Air Force Research
Laboratory Directed Energy Directorate. A portion of the research in this paper was carried out at the Jet Propulsion Laboratory, California Institute of Technology, under a contract with the National Aeronautics and Space Administration. This work was performed [in part] under contract with the California Institute of Technology (Caltech) funded by NASA through the Sagan Fellowship Program. This work was supported by the "Programme National de Plan\'etologie" (PNP) of CNRS/INSU. J.R.G.,
A.S., and M.D.P. were supported in part by the National
Science Foundation Science and Technology Center for
Adaptive Optics, managed by the University of California
at Santa Cruz under cooperative agreement No. AST 98-76783. M.D.P. and R.S. were supported in part by NASA Michelson graduate and postdoctoral fellowships, respectively, under contract to the Jet Propulsion Laboratory (JPL) funded by NASA. JPL is managed for NASA by the California Institute of Technology. R.S. was partially supported by an AMNH Kalbfleisch Fellowship. J.L. was also partially supported by an AMNH Kade Fellowship. The Lyot
Project is also grateful to the Cordelia Corporation, Hilary
and Ethel Lipsitz, the Vincent Astor Fund, Judy
Vale, Alison Cooney, and an anonymous donor, who initiated
the project. This research has made use of the SIMBAD database,
operated at CDS, Strasbourg, France.


\clearpage
\section*{Appendix A}

It was discussed in Sec. \ref{Bayes} that disregarding the $p_\mathrm{j}^*$ in Equation (\ref{CI2}) relaxes the constraint on $f$ and yields a larger CI $\left[0,f_\mathrm{max}\right]$. Looking at $f_\mathrm{max}$ as a function of $\alpha$ and the $p_j$'s, demonstrating the precedent result is equivalent to showing that $\frac{\partial f_\mathrm{max}}{\partial p_\mathrm{j}}|_\alpha(\alpha,p_1,...,p_N)\leqslant 0$ for any $\alpha\in]0,1[$ and $p_1,...,p_N\in\left[0,1\right]^N$. Since the function $ f_\mathrm{max}(\alpha,p_1,...,p_N)$ is an implicit function, let us work first on:
\begin{equation}
\alpha_N(f_\mathrm{max},p_1,...,p_N)=\frac{\int^{f_\mathrm{max}}_0\prod^{N}_{\mathrm{j}=1} (1-fp_\mathrm{j}) df}{\int^1_0\prod^{N}_{\mathrm{j}=1} (1-fp_\mathrm{j}) df}.
\end{equation}
Since $\alpha_N(f_\mathrm{max},p_1,...,p_N)$ is an increasing function of $f_\mathrm{max}$:
\begin{equation}\label{demo1}
\frac{\partial f_\mathrm{max}}{\partial p_\mathrm{j}}\Big|_\alpha\leqslant 0 \Leftrightarrow \frac{\partial \alpha_N}{\partial p_\mathrm{j}}\Big|_{f_\mathrm{max}}=- \frac{\partial \alpha_N}{\partial {f_\mathrm{max}}}\Big|_{p_\mathrm{j}}\cdot \frac{\partial f_\mathrm{max}}{\partial p_\mathrm{j}}\Big|_\alpha \geqslant 0
\end{equation}

In order to demonstrate the second proposition in Equation (\ref{demo1}), we first have to demonstrate that 
\begin{equation}\label{demo2}
\alpha_N(f_\mathrm{max},p_1,...,p_N)>\alpha_{N-1}(f_\mathrm{max},p_1,...,p_{\mathrm{j}-1},p_{\mathrm{j}+1},...,p_{N}).
\end{equation}
Because each $p_\mathrm{j}$ play a similar role, we can always choose $\mathrm{j}=N$ for the demonstration. This result is only a mathematical statement of the fact that when you add a star to the survey, the constraints that you put cannot decrease.
\begin{quotation}
proof: 
\begin{equation}
\alpha_N(0,p_1,...,p_N)=\alpha_{N-1}(0,p_1,...,p_{N-1})=0
\end{equation}
and since $0\leqslant (1-fp_N)<1$:
\begin{eqnarray}
\frac{\partial \alpha_N}{\partial {f}}\Big|_{p_\mathrm{j}}(0,p_1,...,p_{N})&=&\frac{1}{\int^1_0\prod^{N}_{\mathrm{j}=1} (1-fp_\mathrm{j}) df} \nonumber \\
& >&
\frac{1}{\int^1_0\prod^{N-1}_{\mathrm{j}=1} (1-fp_\mathrm{j}) df} \nonumber \\
&>&\frac{\partial \alpha_{N-1}}{\partial {f}}\Big|_{p_\mathrm{j}}(0,p_1,...,p_{N-1}).
\end{eqnarray}
So, $\exists\  \epsilon >0 \  /  \ \alpha_N(\epsilon,p_1,...,p_N)> \alpha_{N-1}(\epsilon,p_1,...,p_{N-1}) $. In order to extend this property for $f\in]0,1[$, we must prove that these two functions do not cross. Since $ \alpha_N(0)=\alpha_{N-1}(0)=0$ and $\alpha_N(1)=\alpha_{N-1}(1)=1$, Rolle's theorem states that if the derivatives of the two functions take the same value only once, the two curves cannot cross. Actually, if
\begin{equation}
\frac{\partial \alpha_N}{\partial f}(f)=\frac{\partial \alpha_{N-1}}{\partial f}(f),
\end{equation}
$f$ must satisfy:
\begin{equation}
f=\frac{\int^{1}_0\prod^{N-1}_{\mathrm{j}=1} x(1-xp_\mathrm{j}) dx}{ \int^1_0\prod^{N-1}_{\mathrm{j}=1} (1-xp_\mathrm{j}) dx},
\end{equation}
and is thus uniquely defined. Therefore we demonstrated that the relation \ref{demo2} holds for $f_\mathrm{max}\in]0,1[$.
\end{quotation}

We can now look at $\frac{\partial \alpha_N}{\partial p_N}|_{f_\mathrm{max}}$:
\begin{eqnarray}
\frac{\partial \alpha_N}{\partial p_N}|_{f_\mathrm{max}}&=&\Bigg[
\int^{f_\mathrm{max}}_0\prod^{N}_{\mathrm{j}=1} (1-fp_\mathrm{j}) df\int^1_0\prod^{N-1}_{\mathrm{j}=1} f(1-fp_\mathrm{j}) df \nonumber \\
&-&\int^1_0\prod^{N}_{\mathrm{j}=1} (1-fp_\mathrm{j}) df\int^{f_\mathrm{max}}_0\prod^{N-1}_{\mathrm{j}=1} f(1-fp_\mathrm{j}) df
\Bigg]  \nonumber \\
&/&(\int^1_0\prod^{N}_{\mathrm{j}=1} (1-fp_\mathrm{j}) df)^2  \nonumber \\
\frac{\partial \alpha_N}{\partial p_N}|_{f_\mathrm{max}}&=&\Bigg[   \int^1_0\prod^{N}_{\mathrm{j}=1} (1-fp_\mathrm{j}) df\int^{f_\mathrm{max}}_0\prod^{N-1}_{\mathrm{j}=1} (1-fp_\mathrm{j}+1)(1-fp_\mathrm{j}) df
 \nonumber \\
&-&\int^{f_\mathrm{max}}_0\prod^{N}_{\mathrm{j}=1} (1-fp_\mathrm{j}) df\int^1_0\prod^{N-1}_{\mathrm{j}=1}(1-fp_\mathrm{j}+1)(1-fp_\mathrm{j}) df
\Bigg]  \nonumber \\
&/&(p_N\int^1_0\prod^{N}_{\mathrm{j}=1} (1-fp_\mathrm{j}) df)^2.
\end{eqnarray}
Manipulating the integrals, we see that 
\begin{eqnarray}
\frac{\partial \alpha_N}{\partial p_N}|_{f_\mathrm{max}}&=&\Bigg[\alpha_N(f_\mathrm{max},p_1,...,p_N)-\alpha_{N-1}(f_\mathrm{max},p_1,...,p_{N-1})\Bigg] \nonumber \\
&\times&\int^1_0\prod^{N-1}_{\mathrm{j}=1} f(1-fp_\mathrm{j}) df \int^1_0\prod^{N}_{\mathrm{j}=1} (1-fp_\mathrm{j}) df  \nonumber \\
&/&(p_N\int^1_0\prod^{N}_{\mathrm{j}=1} (1-fp_\mathrm{j}) df)^2,
\end{eqnarray}
which is obviously positive considering Equation (\ref{demo2}). This concludes to demonstrate that
\begin{equation}
\frac{\partial f_\mathrm{max}}{\partial p_\mathrm{j}}|_\alpha(\alpha,p_1,...,p_N)\leqslant 0
\end{equation}
and our result.

\bibliography{biblio} 
\bibliographystyle{apj} 




 \end{document}